# Double Antisymmetry and the Rotation-Reversal Space Groups

Brian K. VanLeeuwen[a]*, Venkatraman Gopalan[a]* and Daniel B. Litvin[b]*

[a]Department of Materials Science and Engineering, The Pennsylvania State University, University Park, Pennsylvania, 16803, USA, and [b]Department of Physics, The Eberly College of Science, The Pennsylvania State University, Penn State Berks, P.O. Box 7009, Reading, Pennsylvania, 19610, USA

Correspondence email: Brian.VanLeeuwen@gmail.com; vxg8@psu.edu; u3c@psu.edu

**Synopsis**

It was found that there are 17,803 types of double antisymmetry groups. This is four fewer than previously thought. When rotation-reversal symmetry and time-reversal symmetry are considered, this implies that there are 17,803 distinct types of symmetry a crystal may exhibit.

**Abstract**

Rotation-reversal symmetry was recently introduced to generalize the symmetry classification of rigid static rotations in crystals such as tilted octahedra in perovskite structures and tilted tetrahedral in silica structures. This operation has important implications for crystallographic group theory, namely that new symmetry groups are necessary to properly describe observations of rotation-reversal symmetry in crystals. When both rotation-reversal symmetry and time-reversal symmetry are considered in conjunction with space group symmetry, it is found that there are 17,803 types of symmetry, called double antisymmetry, which a crystal structure can exhibit. These symmetry groups have the potential to advance understanding of polyhedral rotations in crystals, the magnetic structure of crystals, and the coupling thereof. The full listing of the double antisymmetry space groups can be found in the supplemental materials of the present work and online at http://sites.psu.edu/gopalan/research/symmetry/

# 1. Introduction

Rotation-reversal symmetry (Gopalan & Litvin, 2011) was introduced to generalize the symmetry classification of tilted octahedra perovskite structures (Glazer, 2011). The rotation-reversal operation, represented by 1* (previously represented by $1^\Phi$) was compared by analogy to the well-known time-reversal operation, represented by 1' (see Figure 1). Although reversing time in a crystal is not something which can performed experimentally, it is nonetheless useful for describing the magnetic symmetry of a crystal structure and this magnetic symmetry description has important consequences which can be observed by experiment (Opechowski, 1986). Likewise, rotation-reversal symmetry is useful in describing the symmetry of a crystal structure composed of molecules or polyhedral units. For example, the structure conventionally described in Glazer notation (Glazer, 1972) as $a_o^+ a_o^+ c_o^+$ and classified with the group Immm1' is now classified with the rotation-reversal group I4*/mmm1'. A $a_o^+ a_o^+ c_o^+$ structure with a spin along the z-direction in each octahedron was formerly classified with the group Im'm'm is now classified with the group I4*/mm'm'*. The rotation-reversal space groups used in this new classification of tilted octahedra perovskites are isomorphic to, i.e. have the same abstract mathematical structure, as *double antisymmetry* space groups (Zamorzaev, 1976). As rotation-reversal space groups are isomorphic to double antisymmetry space groups, in the present work we will use the terminology and notation associated with double antisymmetry space groups.

Double antisymmetry space groups are among the generalizations of the crystallographic groups which began with Heesch (1930) and Shubnikov (1951) and continued to include a myriad of generalizations under various names as antisymmetry groups, cryptosymmetry groups, quasisymmetry groups, color groups and metacrystallographic groups (see reviews by Koptsik (1967), Zamorzaev & Palistrant (1980), Opechowski (1986), and Zamorzaev (1988)). Only some of these groups have been explicitly listed, e.g. the black and white space groups (Belov et al., 1955, 1957) and various multiple antisymmetry (Zamorzaev, 1976) and color groups (Zamorzaev et al., 1978). While no explicit listing of the double antisymmetry space groups has been given, the number of these groups, and other generalizations of the crystallographic groups, has been calculated (see (Zamorzaev, 1976, 1988; Zamorzaev & Palistrant, 1980; Jablan, 1987, 1990, 1992, 1993a; b, 2002; Palistrant & Jablan, 1991; Radovic & Jablan, 2005)).

In Section 2, we shall define double antisymmetry space groups and specify which of these groups we shall explicitly tabulate. This is followed by the details of the procedure used in their tabulation. In Section 3, we set out the format of the tables listing these groups. Section 4 describes example diagrams of double antisymmetry space groups (**Figure 4a:** No. 8543 C2'/m*, **Figure 4b:** No.

16490 I4*/mm'm'*, and **Figure 4c:** No. 13461 Ib'*c'a'; colors seen only online).  Section 5 gives the computational details of how the types were derived.

Zamorzaev & Palistrant (1980) have calculated not only the total number of types of double antisymmetry space groups, but they have also specified the number in sub-categories.  We have found errors in these numbers.  Consequently, the total number of types of groups is different than calculated by Zamorzaev & Palistrant (1980).  This is discussed in Section 6.

## 2. Double Antisymmetry Space Groups

Space groups, in the present work, will be limited to the conventional three-dimensional crystallographic space groups as defined in the International Tables for Crystallography (2006).  An *antisymmetry space group* is similar to a space group, but some of the symmetry elements may also "flip" space between two possible states, e.g. (**r**,*t*) and (**r**,-*t*).  A *double antisymmetry space group* extends this concept to allow symmetry elements to flip space in two independent ways between four possible states, e.g. (**r**,*t*,$\phi$), (**r**,-*t*,$\phi$), (**r**,*t*,-$\phi$), and (**r**,-*t*,-$\phi$).  A more precise definition will be given below.

To precisely define antisymmetry space groups, we will start by defining an anti-identity.  An operation, e.g. 1', is an anti-identity if it has the following properties:

- Self-inverse: 1'·1' = 1 where 1 is identity

- Commutivity: 1'·g = g·1' for all elements g of **E**(3)

- 1' is not an element of **E**(3)

**E**(3) is the 3-dimensional Euclidean group, i.e. the group of all distance-preserving transformations of 3-dimensional Euclidean space.  A *space group* can be defined as a group **G** ⊂ **E**(3) such that the subgroup composed of all translations in **G** is minimally generated by a set of three translations.  We can extend this to define antisymmetry space groups as follows:

> Let an *n-tuple antisymmetry space group* be defined as a group **G** ⊂ **E**(3) × **P** such that the subgroup composed of all translations in **G** is minimally generated by a set of three translations and **P** is minimally generated by a set of *n* anti-identities and is isomorphic to $\mathbb{Z}_2^n$.

Thus, for single antisymmetry space groups, **P** is generated by just one anti-identity, for double, two, for triple, three, and so forth.  It should be noted that the above definition could be generalized to arbitrary spaces and coloring schemes by changing **E**(3) and **P** respectively, but that is beyond the scope of the present work.

Let the two anti-identities which generate **P** for double antisymmetry space groups be labeled as 1' and 1*. The product of 1' and 1* is also an anti-identity which will be labeled 1'*. The coloring of 1', 1*, and 1'* (colors seen online only) is intended to assist the reader and has no special meaning beyond that. Double antisymmetry has a total of three anti-identities: 1', 1*, and 1'*. Note that these three anti-identities are not independent because each can be generated from the product of the other two. So although we have three anti-identities, only two are independent and thus we call it "double antisymmetry" (more generally $n$-antisymmetry has $2^n$-1 anti-identities). 1' generates the group **1'** = {1, 1'}, 1* generates the group **1*** = {1, 1*}, and together 1' and 1* generate the group **1'1*** = {1, 1', 1*, 1'*}. For double antisymmetry space groups, **P** = **1'1***.

Figure 2 shows how the elements of **1'1*** multiply. To evaluate the product of two elements of **1'1*** with the multiplication table given in Figure 2a, we find the row associated with the first element and the column associated with second element, e.g. for 1'·1*, go to the second row third column to find 1'*. To evaluate the product of two elements of **1'1*** with the Cayley graph given in Figure 2b, we start from the circle representing the first element and follow the arrow representing the second, e.g. for 1'·1*, we start on the red circle (1') and take the blue path (1*) to the green circle (1'*).

### 2.1. The structure of double antisymmetry groups

When a spatial transformation is coupled with an anti-identity, we shall say it is *colored* with that anti-identity. This is represented by adding ', *, or '* to the end of the symbol representing the spatial transformation, e.g. a four-fold rotation coupled with time-reversal (i.e. the product of 4 and 1') is 4'.

We shall say that all double antisymmetry groups can be constructed by coloring the elements of a *colorblind parent group*. In the case of a double antisymmetry space group, the colorblind parent group, **Q**, is one of the crystallographic space groups. There are four different ways of coloring an element of **Q**, namely coloring with 1, 1', 1*, or 1'* which we shall then refer to as being *colorless*, *primed*, *starred*, or *prime-starred* respectively. Let **Q1'1*** be the group formed by including all possible colorings of the elements of **Q**, i.e. the direct product of **Q** and **1'1***. Since **Q1'1*** contains all possible colorings of the elements of **Q**, every double antisymmetry group whose colorblind parent is **Q** must be a subgroup of **Q1'1***.

Every subgroup of **Q1'1*** whose colorblind parent is **Q** is of the form of one of the twelve categories of double antisymmetry groups listed in Table 1. The formulae of Table 1 are represented visually in Appendix A using Venn diagrams.

## 2.2. Example of generating double antisymmetry groups

As an example, consider applying the formulae in Table 1 to point group **222**. **222** has four elements: $\{1,2_x,2_y,2_z\}$. **222** has three index-2 subgroups: $\{1,2_x\}$, $\{1,2_y\}$, and $\{1,2_z\}$ which will be denoted by $\mathbf{2_x}$, $\mathbf{2_y}$, and $\mathbf{2_z}$ respectively. The subscripts indicate the axes of rotation. Applying the formulae in Table 1 yields the following:

Category 1): **Q**

1. $\mathbf{Q = 222} \rightarrow \mathbf{Q} = \{1,2_x,2_y,2_z\}$

Category 2): **Q + Q1'**

2. $\mathbf{Q = 222} \rightarrow \mathbf{Q1'} = \{1,2_x,2_y,2_z,1',2_x',2_y',2_z'\}$

Category 3): **H + (Q – H)1'**

3. $\mathbf{Q = 222}, \mathbf{H = 2_x} \rightarrow \mathbf{Q(H)} = \{1,2_x,2_y',2_z'\}$

4. $\mathbf{Q = 222}, \mathbf{H = 2_y} \rightarrow \mathbf{Q(H)} = \{1,2_x',2_y,2_z'\}$

5. $\mathbf{Q = 222}, \mathbf{H = 2_z} \rightarrow \mathbf{Q(H)} = \{1,2_x',2_y',2_z\}$

Category 4): **Q + Q1***

6. $\mathbf{Q = 222} \rightarrow \mathbf{Q1^*} = \{1,2_x,2_y,2_z,1^*,2_x^*,2_y^*,2_z^*\}$

Category 5): **Q + Q1' + Q1* + Q1'***

7. $\mathbf{Q = 222} \rightarrow \mathbf{Q1'1^*} = \{1,2_x,2_y,2_z,1',2_x',2_y',2_z',1^*,2_x^*,2_y^*,2_z^*,1'^*,2_x'^*,2_y'^*,2_z'^*\}$

Category 6): **H + (Q-H)1' + H1* + (Q-H)1'***

8. $\mathbf{Q = 222}, \mathbf{H = 2_x} \rightarrow \mathbf{Q(H)1^*} = \{1,2_x,2_y',2_z',1^*,2_x^*,2_y'^*,2_z'^*\}$

9. $\mathbf{Q = 222}, \mathbf{H = 2_y} \rightarrow \mathbf{Q(H)1^*} = \{1,2_x',2_y,2_z',1^*,2_x'^*,2_y^*,2_z'^*\}$

10. $\mathbf{Q = 222}, \mathbf{H = 2_z} \rightarrow \mathbf{Q(H)1^*} = \{1,2_x',2_y',2_z,1^*,2_x'^*,2_y'^*,2_z^*\}$

Category 7): **H + (Q-H)1***

11. $\mathbf{Q = 222}, \mathbf{H = 2_x} \rightarrow \mathbf{Q\{H\}} = \{1,2_x,2_y^*,2_z^*\}$

12. $\mathbf{Q = 222}, \mathbf{H = 2_y} \rightarrow \mathbf{Q\{H\}} = \{1,2_x^*,2_y,2_z^*\}$

13. $\mathbf{Q = 222}, \mathbf{H = 2_z} \rightarrow \mathbf{Q\{H\}} = \{1,2_x^*,2_y^*,2_z\}$

Category 8): **Q + Q1'***

14. $\mathbf{Q = 222} \rightarrow \mathbf{Q1'^*} = \{1,2_x,2_y,2_z,1'^*,2_x'^*,2_y'^*,2_z'^*\}$

Category 9): **H** + (**Q**-**H**)1* + **H**1' + (**Q**-**H**)1'*

    15. $Q = \mathbf{222}, H = \mathbf{2_x} \rightarrow Q\{H\}\mathbf{1'} = \{1, 2_x, 2_y{}^*, 2_z{}^*, 1', 2_x{}', 2_y{}'^*, 2_z{}'^*\}$

    16. $Q = \mathbf{222}, H = \mathbf{2_y} \rightarrow Q\{H\}\mathbf{1'} = \{1, 2_x{}^*, 2_y, 2_z{}^*, 1', 2_x{}'^*, 2_y{}', 2_z{}'^*\}$

    17. $Q = \mathbf{222}, H = \mathbf{2_z} \rightarrow Q\{H\}\mathbf{1'} = \{1, 2_x{}^*, 2_y{}^*, 2_z, 1', 2_x{}'^*, 2_y{}'^*, 2_z{}'\}$

Category 10): **H** + (**Q**-**H**)1' + **H**1'* + (**Q**-**H**)1*

    18. $Q = \mathbf{222}, H = \mathbf{2_x} \rightarrow Q(H)\mathbf{1'}^* = \{1, 2_x, 2_y{}', 2_z{}', 1'^*, 2_x{}'^*, 2_y{}^*, 2_z{}^*\}$

    19. $Q = \mathbf{222}, H = \mathbf{2_y} \rightarrow Q(H)\mathbf{1'}^* = \{1, 2_x{}', 2_y, 2_z{}', 1'^*, 2_x{}^*, 2_y{}'^*, 2_z{}^*\}$

    20. $Q = \mathbf{222}, H = \mathbf{2_z} \rightarrow Q(H)\mathbf{1'}^* = \{1, 2_x{}', 2_y{}', 2_z, 1'^*, 2_x{}^*, 2_y{}^*, 2_z{}'^*\}$

Category 11): **H** + (**Q**-**H**)1'*

    21. $Q = \mathbf{222}, H = \mathbf{2_x} \rightarrow Q(H)\{H\} = \{1, 2_x, 2_y{}'^*, 2_z{}'^*\}$

    22. $Q = \mathbf{222}, H = \mathbf{2_y} \rightarrow Q(H)\{H\} = \{1, 2_x{}'^*, 2_y, 2_z{}'^*\}$

    23. $Q = \mathbf{222}, H = \mathbf{2_z} \rightarrow Q(H)\{H\} = \{1, 2_x{}'^*, 2_y{}'^*, 2_z\}$

Category 12): **H**∩**K** + (**H**-**K**)1* + (**K**-**H**)1' + (**Q**-(**H**+**K**))1'*

    24. $Q = \mathbf{222}, H = \mathbf{2_y}, K = \mathbf{2_z} \rightarrow Q(H)\{K\} = \{1, 2_x{}'^*, 2_y{}^*, 2_z{}'\}$

    25. $Q = \mathbf{222}, H = \mathbf{2_x}, K = \mathbf{2_z} \rightarrow Q(H)\{K\} = \{1, 2_x{}^*, 2_y{}'^*, 2_z{}'\}$

    26. $Q = \mathbf{222}, H = \mathbf{2_x}, K = \mathbf{2_y} \rightarrow Q(H)\{K\} = \{1, 2_x{}^*, 2_y{}', 2_z{}'^*\}$

    27. $Q = \mathbf{222}, H = \mathbf{2_z}, K = \mathbf{2_y} \rightarrow Q(H)\{K\} = \{1, 2_x{}'^*, 2_y{}', 2_z{}^*\}$

    28. $Q = \mathbf{222}, H = \mathbf{2_z}, K = \mathbf{2_x} \rightarrow Q(H)\{K\} = \{1, 2_x{}', 2_y{}'^*, 2_z{}^*\}$

    29. $Q = \mathbf{222}, H = \mathbf{2_y}, K = \mathbf{2_x} \rightarrow Q(H)\{K\} = \{1, 2_x{}', 2_y{}^*, 2_z{}'^*\}$

Note that although 29 double antisymmetry point groups are generated from using **222** as a colorblind parent group, they are not all of distinct antisymmetry point group types, as is explained in Section 2.3. In the above example, there are only 12 unique types of groups, as discussed further on. Two additional examples, point group **2/m** and space group **Cc**, are given in Appendix B.

    Since the index-2 subgroups of the crystallographic space groups are already known and available in the International Tables for Crystallography, applying this set of formulae is straightforward. If applied to a representative group of each of the 230 crystallographic space group types, 38,290 double antisymmetry space groups are generated. These 38,290 generated groups can be sorted into 17,803 equivalence classes, i.e. *double antisymmetry group types*, by applying an equivalence relation.

## 2.3. Double antisymmetry group types and the proper affine equivalence relation

The well-known 230 *crystallographic space group types* given in the International Tables for Crystallography are the *proper affine classes of space groups* ("types" is used instead of "classes" to avoid confusion with "crystal classes" (2006)). The equivalence relation of proper affine classes is as follows: two space groups are equivalent if and only if they can be bijectively mapped by a proper affine transformation (*proper* means *chirality-preserving*) (Opechowski, 1986). In the literature, the "*space group types*" are often referred to as simply "*space groups*" when the distinction is unnecessary.

For the present work, we will use "double antisymmetry group types" to refer to the proper affine classes of double antisymmetry groups. This is consistent with Zamorzaev's works on generalized antisymmetry (Zamorzaev, 1976, 1988; Zamorzaev et al., 1978; Zamorzaev & Palistrant, 1980).

As an example, we consider the proper affine equivalence classes of the twenty-nine $Q = 222$ double antisymmetry groups generated using the formulae given in Table 1: Only twelve such classes exist, one in each category. For categories 1), 2), 4), 5), and 8), the reason for this is that there is only one group generated in each to begin with. For categories 3), 6), 7), 9), 10), and 11), there are three groups generated which are related to each other by 120 degree rotations (e.g. $\{1, 2_x, 2_y', 2_z'\} = 3_{xyz} \cdot \{1, 2_x', 2_y, 2_z'\} \cdot 3_{xyz}^{-1} = 3_{xyz}^{-1} \cdot \{1, 2_x', 2_y', 2_z\} \cdot 3_{xyz}$) and therefore they are members of the same equivalence class. For category 12), the six generated groups are all in the same equivalence class because $\{1, 2_x'^*, 2_y^*, 2_z'\} = 3_{xyz} \cdot \{1, 2_x^*, 2_y', 2_z'^*\} \cdot 3_{xyz}^{-1} = 3_{xyz}^{-1} \cdot \{1, 2_x', 2_y'^*, 2_z^*\} \cdot 3_{xyz} = 4_x \cdot \{1, 2_x'^*, 2_y', 2_z^*\} \cdot 4_x^{-1} = 4_x \cdot 3_{xyz} \cdot \{1, 2_x', 2_y^*, 2_z'^*\} \cdot 3_{xyz}^{-1} \cdot 4_x^{-1} = 4_x \cdot 3_{xyz}^{-1} \cdot \{1, 2_x^*, 2_y'^*, 2_z'\} \cdot 3_{xyz} \cdot 4_x^{-1}$. This is demonstrated with point group diagrams in Figure 3.

Thus the proper affine equivalence classes of the $Q = 222$ double antisymmetry point groups can be represented by:

Category 1): **Q**

1. $Q = 222 \rightarrow Q = \{1, 2_x, 2_y, 2_z\}$

Category 2): **Q + Q1'**

2. $Q = 222 \rightarrow Q1' = \{1, 2_x, 2_y, 2_z, 1', 2_x', 2_y', 2_z'\}$

Category 3): **H + (Q – H)1'**

3. $Q = 222, H = 2_x \rightarrow Q(H) = \{1, 2_x, 2_y', 2_z'\}$

Category 4): **Q + Q1\***

4. $Q = 222 \rightarrow Q1^* = \{1, 2_x, 2_y, 2_z, 1^*, 2_x^*, 2_y^*, 2_z^*\}$

Category 5): $Q + Q1' + Q1^* + Q1'^*$

5. $Q = 222 \rightarrow Q1'1^* = \{1, 2_x, 2_y, 2_z, 1', 2_x', 2_y', 2_z', 1^*, 2_x^*, 2_y^*, 2_z^*, 1'^*, 2_x'^*, 2_y'^*, 2_z'^*\}$

Category 6): $H + (Q-H)1' + H1^* + (Q-H)1'^*$

6. $Q = 222, H = 2_x \rightarrow Q(H)1^* = \{1, 2_x, 2_y', 2_z', 1^*, 2_x^*, 2_y'^*, 2_z'^*\}$

Category 7): $H + (Q-H)1^*$

7. $Q = 222, H = 2_x \rightarrow Q\{H\} = \{1, 2_x, 2_y^*, 2_z^*\}$

Category 8): $Q + Q1'^*$

8. $Q = 222 \rightarrow Q1'^* = \{1, 2_x, 2_y, 2_z, 1'^*, 2_x'^*, 2_y'^*, 2_z'^*\}$

Category 9): $H + (Q-H)1^* + H1' + (Q-H)1'^*$

9. $Q = 222, H = 2_x \rightarrow Q\{H\}1' = \{1, 2_x, 2_y^*, 2_z^*, 1', 2_x', 2_y'^*, 2_z'^*\}$

Category 10): $H + (Q-H)1' + H1'^* + (Q-H)1^*$

10. $Q = 222, H = 2_x \rightarrow Q(H)1'^* = \{1, 2_x, 2_y', 2_z', 1^*, 2_x'^*, 2_y^*, 2_z^*\}$

Category 11): $H + (Q-H)1'^*$

11. $Q = 222, H = 2_x \rightarrow Q(H)\{H\} = \{1, 2_x, 2_y'^*, 2_z'^*\}$

Category 12): $H \cap K + (H-K)1^* + (K-H)1' + (Q-(H+K))1'^*$

12. $Q = 222, H = 2_y, K = 2_z \rightarrow Q(H)\{K\} = \{1, 2_x'^*, 2_y^*, 2_z'\}$

Proper affine equivalence and other definitions of equivalence are discussed in Appendix C.

### 2.4. Derivation of the double antisymmetry space group types

Double antisymmetry space groups of categories 1) through 11) of Table 1 are already known or easily derived: The group types of category 1) are the well-known 230 conventional space group (2006). The groups of categories 2), 4), 5), and 8) are effectively just products of the groups of category 1) with **1'**, **1\***, **1'1\***, and **1'\*** respectively. The groups of category 3) are the well-known black-and-white space groups (Belov et al., 1955, 1957) (also known as type **M** magnetic space groups (Opechowski, 1986)). The groups of categories 7) and 11) are derived by substituting starred operations and prime-starred operations, respectively, for the primed operations of category 3) groups. And finally, the groups of category 6), 9), and 10) are products of the groups of categories 3), 7), and 11) respectively with **1\***, **1'**, and **1'\*** respectively.

For the groups of category 12) we have used the following four step procedure:

1. For one representative group **Q** from each of the 230 types of crystallographic space groups, we list all subgroups of index two (2006; Aroyo et al., 2006a; b, 2011).

2. We construct and list all double antisymmetry space groups **Q(H){K}** for each representative group **Q** and pairs of distinct subgroups, **H** and **K**, of index two. This step results in 26,052 **Q(H){K}** groups.

3. For every pair of groups **Q₁(H₁){K₁}** and **Q₂(H₂){K₂}** where **Q₁** and **Q₂**, **H₁** and **H₂**, and **K₁** and **K₂**, are pair-wise of the same space group type, we evaluate the proper affine equivalence relation to determine if **Q₁(H₁){K₁}** and **Q₂(H₂){K₂}** are of the same double antisymmetry space group type.

4. From each set of groups belonging to the same double antisymmetry type, we list one representative double antisymmetry space group **Q(H){K}**.

Further details for each of these steps are given in Section 5.

## 3. Tables of Double Antisymmetry Space Groups

### 3.1. Double antisymmetry space group types

The serial numbers and symbols alone are given in "Double Antisymmetry Space Groups Symbols.pdf" [†]. The double antisymmetry space group symbols are based on the Hermann-Mauguin symbol of the colorblind parent space group, e.g. C2'/m* is based on C2/m.

The first part of the symbol gives the lattice (or more precisely the translational subgroup). If there are no colored translations in the group, the symbol is given as P (primitive), C (c-centering), A (a-centering), I (body-centering), F (face-centering), and R (rhombohedral-centering). If there are colored translations, the P, C, A, I, F, and R are followed by three color operations in parenthesis, e.g. C(1,1'*,1')2/m'*. These three color operations denote the coloring of a minimal set of generating lattice translations indicated in Table 3. For example, consider C(1,1'*,1')2/m'*: 1 is in the first position, 1'* is in the second position, and 1' is in the third position. Looking up the lattice symbol "C" in the first column of Table 3, we find that the first, second, and third positions correspond to $t_{[100]}$, $t_{[001]}$, and $t_{[½½0]}$ respectively. The translations of C(1,1'*,1')2/m'* are therefore generated by $t_{[100]}$, $t_{[001]}$'*, and $t_{[½½0]}$'.

The second part of the symbol gives the remaining generators for the double antisymmetry space group. This is also based on the corresponding part of the Hermann-Mauguin symbol of the colorblind parent space group. The Seitz notation of each character is given in "secondPartOfSymbolGenerators.pdf"[†].

Finally, if the group is a member of category 2), 4), 5), 6), 8), 9), or 10), then 1', 1*, 1'1*, 1*, 1'*, 1', 1'*, or 1'* respectively is appended to the end of the symbol.

### 3.2. Using the Computable Document Format (CDF) file

The Computable Document Format (CDF) file, "Double antisymmetry space groups.cdf"[†] provides an interactive way to find the symbols and operations of double antisymmetry space groups. The file is opened with the Wolfram CDF Player which can be downloaded from http://www.wolfram.com/cdf-player/. After opening the file, click "Enable Dynamics" if prompted. Provide the necessary input with the drop down menus.

A tutorial with screenshots is given in "Double antisymmetry space groups CDF tutorial.pdf"[†].

### 3.3. Using the PDF file

In the PDF file, "Double Antisymmetry Space Groups Symbols and Operations.pdf"[†], the 17,803 double antisymmetry space groups are listed sequentially in Seitz notation. The first line of each entry gives the sequential serial number (1 through 17,803) and the double antisymmetry space group symbol. The remaining listing gives, in Seitz notation, a set of coset representatives of the double antisymmetry group with respect to its translational subgroup. Three examples from the listings of representative double antisymmetry space groups are given in Table 2.

### 3.4. Machine-readable file

A "machine-readable" file, "DASGMachineReadable.txt"[†] is intended to provide a simple way to use the double antisymmetry space groups in code or software such as MatLab or Mathematica. The structure of the file is given in Appendix D. The file, "Import DASGMachineReadable.nb"[†] has been provided to facilitate loading into Mathematica.

## 4. Symmetry Diagrams

Symmetry diagrams have been made for the example double antisymmetry space groups listed in Table 2. These diagrams are intended to extend the conventional space group diagrams such as those in the International Tables for Crystallography.

In **Figure 4a**, double antisymmetry space group No. 8543, C2'/m*, is projected along the *b*-axis. The red lenticular shape with the green dot inset, i.e. 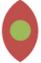, indicates the position of a primed two-fold rotation axis (2') and a prime-starred point of inversion (1'*) coinciding with the plane of projection. Likewise, the lenticular shapes with "tails", i.e. 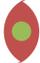, indicate screw axes. The same symbols but with a "¼" placed near the upper left corner indicate that the prime-starred point of inversion is out-plane by a displacement of ¼ along the *b*-axis. The 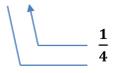 on the bottom left of the diagram indicates that there is a starred mirror (m*) coinciding with the plane of projection and a starred glide (a*) out-plane by a displacement of ¼ along the *b*-axis.

In **Figure 4b**, double antisymmetry space group No. 16490, I4*/mm'm'*, is projected along the *c*-axis. In **Figure 4c**, double antisymmetry space group No. 13461, Ib'*c'a', is projected along the *c*-axis. As with **Figure 4a**, the symbols in these diagrams are naturally extended from those used for conventional space groups in the International Tables for Crystallography.

## 5. Computation Details

The majority of the computation was performed in Mathematica using faithful matrix representations of space groups. These matrices were downloaded from the Bilbao crystallographic server (Aroyo et al., 2006a; b, 2011). These representations correspond to the standard setting for each of the 230 space group types as given in the International Tables for Crystallography.

### 5.1. The faithful matrix representation of space groups

Every space group operation can be broken up into a linear transformation $R$ and a translation $t$ which transform the Cartesian coordinates $(r_1, r_2, r_3)$ into $(r_1', r_2', r_3')$:

**Equation 1**

$$\mathbf{r'} = R\mathbf{r} + \mathbf{t}$$

$$\begin{pmatrix} r_1' \\ r_2' \\ r_3' \end{pmatrix} = \begin{pmatrix} R_{11} & R_{12} & R_{13} \\ R_{21} & R_{22} & R_{23} \\ R_{31} & R_{32} & R_{33} \end{pmatrix} \begin{pmatrix} r_1 \\ r_2 \\ r_3 \end{pmatrix} + \begin{pmatrix} t_1 \\ t_2 \\ t_3 \end{pmatrix}$$

It is convenient to condense this linear transformation and translation into a single square matrix *a*:

**Equation 2**

$$\mathbf{r'} = a\mathbf{r}$$

$$\begin{pmatrix} r_1' \\ r_2' \\ r_3' \\ 1 \end{pmatrix} = \begin{pmatrix} R_{11} & R_{12} & R_{13} & t_1 \\ R_{21} & R_{22} & R_{23} & t_2 \\ R_{31} & R_{32} & R_{33} & t_3 \\ 0 & 0 & 0 & 1 \end{pmatrix} \begin{pmatrix} r_1 \\ r_2 \\ r_3 \\ 1 \end{pmatrix}$$

Note that the final row is necessary to make a square 4x4 matrix but contains no information specific to the transformation. Under this representation, the group operation is evaluated by matrix multiplication, i.e. $g_1g_2$ is performed by multiplying the matrices representing $g_1$ and $g_2$ to get a product matrix which is also a member of the matrix representation of the group. The representation of the inverse of a group element is the matrix inverse of the representation thereof. The sign of the determinant of the matrix representation of an operation determines if the operation is proper (orientation-preserving). A positive determinant means that it is proper.

For a set of *n*-by-*n* matrices $S$ and an *n*-by-*n* matrix $A$, we will denote the set formed by the similarity transformation of each element of $S$ by $A$ as $ASA^{-1}$, i.e. $ASA^{-1} = \{AsA^{-1} : s \in S\}$.

### 5.2. Index-2 subgroups

To evaluate the formulae in Table 1 for any given colorblind parent space group $Q$, the index-2 subgroups of $Q$ are needed. The index-2 subgroups of a space group are also space groups themselves. As such, every index-2 subgroups of space group $Q$ must be one of the 230 types of space groups. Using this, we can specify an index-2 subgroup of $Q$ by specifying the type of the subgroup (1 to 230) and a transformation from a standard representative group of that type, i.e. $H$, a subgroup of $Q$, can be specified by a standard representative group $H_0$ and the transformation, $T$ such that $H = TH_0T^{-1}$.

The index-2 subgroup data was downloaded from the Bilbao Crystallographic Server. All together there are 1,848 index-2 subgroups among the 230 representative space groups. Each entry (out of 1,848) consisted of: a number between 1 and 230 for $Q$, a number between 1 and 230 for $H_0$, and a 4-by-4 matrix for $T$.

### 5.3. Normalizer method for evaluating the proper affine equivalence of Q(H){K} groups

The proper affine equivalence relation can be defined as: two groups, $G_1$ and $G_2$, are equivalent if and only if $G_1$ can be bijectively mapped to $G_2$ by a proper affine transformation, $a$. Expressed in mathematical short-hand, this is:

**Equation 3**

$$G_2 \sim G_1 \equiv \exists a \in \mathcal{A}^+ : (G_2 = aG_1a^{-1})$$

where $\sim$ is the proper affine equivalence relational operator, $\equiv$ is logical equivalence ("is logically equivalent to"), $\exists a$ is the existential quantification of $a$ ("there exists $a$"), $\in$ means "an element of", : means "such that", and $\mathcal{A}^+$ is the group of proper affine transformations.

For evaluating the proper affine equivalence of a pair of **Q(H){K}** groups, if the space group types of **Q**, **H**, and **K** of one of the **Q(H){K}** groups are not the same as those of the other, then the proper affine equivalence relation fails and no further work is necessary. For a pair of **Q(H){K}** groups where they are the same, we have derived a method to evaluate proper affine equivalence based on space group normalizer groups. To begin this derivation, we can expand the proper affine equivalence for **Q(H){K}** groups to:

**Equation 4**

$$Q(H_2)\{K_2\} \sim Q(H_1)\{K_1\} \equiv \exists a \in \mathcal{A}^+ : (Q = aQa^{-1}) \wedge (H_2 = aH_1a^{-1}) \wedge (K_2 = aK_1a^{-1})$$

where $\wedge$ denotes logical conjuction ("and"). We can use the definition of a normalizer group, i.e. $N_G(S) = \{g \in G : S = gSg^{-1}\}$ (Opechowski, 1986), to get:

**Equation 5**

$$Q(H_2)\{K_2\} \sim Q(H_1)\{K_1\} \equiv \exists a \in N_{\mathcal{A}^+}(Q) : (H_2 = aH_1a^{-1}) \wedge (K_2 = aK_1a^{-1})$$

where $N_{\mathcal{A}^+}(Q)$ is the proper affine normalizer group of $Q$, i.e. $N_{\mathcal{A}^+}(Q) = \{a \in \mathcal{A}^+ : Q = aQa^{-1}\}$.

$H_1$ and $H_2$ are mapped by proper affine transformations, $T_{H_1}$ and $T_{H_2}$ respectively, from a standard representation, $H_0$, as follows: $H_1 = T_{H_1} H_0 T_{H_1}^{-1}$ and $H_2 = T_{H_2} H_0 T_{H_2}^{-1}$. Likewise, $K_1$ and $K_2$ are mapped by proper affine transformations, $T_{K_1}$ and $T_{K_2}$ respectively, from a standard representation, $K_0$, as follows: $K_1 = T_{K_1} K_0 T_{K_1}^{-1}$ and $K_2 = T_{K_2} K_0 T_{K_2}^{-1}$. Thus, by substitution, $(H_2 = aH_1a^{-1}) \wedge$

$(K_2 = aK_1a^{-1})$ is equivalent to $(T_{H_2}H_0T_{H_2}^{-1} = aT_{H_1}H_0T_{H_1}^{-1}a^{-1}) \wedge (T_{K_2}K_0T_{K_2}^{-1} = aT_{K_1}K_0T_{K_1}^{-1}a^{-1})$ which can be rearranged to $\left(H_0 = (T_{H_2}^{-1}aT_{H_1})H_0(T_{H_2}^{-1}aT_{H_1})^{-1}\right) \wedge \left(K_0 = (T_{K_2}^{-1}aT_{K_1})K_0(T_{K_2}^{-1}aT_{K_1})^{-1}\right)$. By applying the definition of a normalizer group again, we find that $T_{H_2}^{-1}aT_{H_1} \in N_{\mathcal{A}^+}(H_0)$ and $T_{K_2}^{-1}aT_{K_1} \in N_{\mathcal{A}^+}(K_0)$, which rearrange to $a \in T_{H_2}N_{\mathcal{A}^+}(H_0)T_{H_1}^{-1}$ and $a \in T_{K_2}N_{\mathcal{A}^+}(K_0)T_{K_1}^{-1}$ respectively. Therefore, the proper affine equivalence of $Q(H_1)\{K_1\}$ and $Q(H_2)\{K_2\}$ is logically equivalent to the existence of a non-empty intersection of $N_{\mathcal{A}^+}(Q)$, $T_{H_2}N_{\mathcal{A}^+}(H_0)T_{H_1}^{-1}$, and $T_{K_2}N_{\mathcal{A}^+}(K_0)T_{K_1}^{-1}$:

**Equation 6**

$$Q(H_1)\{K_1\} \sim Q(H_2)\{K_2\} \equiv N_{\mathcal{A}^+}(Q) \cap T_{H_2}N_{\mathcal{A}^+}(H_0)T_{H_1}^{-1} \cap T_{K_2}N_{\mathcal{A}^+}(K_0)T_{K_1}^{-1} \neq \emptyset$$

This simplifies the problem of evaluating the equivalence relation to either proving that the intersection has at least one member or proving that it does not. To do this, we applied Mathematica's built-in "FindInstance" function. As with the subgroup data, the normalizer group data was downloaded from the Bilbao Crystallographic Server. As previously discussed, the formulae in Table 1 generate 38,290 double antisymmetry space groups when applied to all 230 representative space groups. With the aid of Mathematica, these 38,290 double antisymmetry space groups were partitioned by the equivalence relation given in Equation 6 into 17,803 proper affine equivalence classes, i.e. 17,803 *double antisymmetry group types*.

This method can be easily generalized to other types of antisymmetry and color symmetry. For example, for antisymmtery groups formed from one index-2 subgroup, such as $Q(H)$ groups, the condition simply reduces to the following:

**Equation 7**

$$Q(H_1) \sim Q(H_2) \equiv N_{\mathcal{A}^+}(Q) \cap T_{H_2}N_{\mathcal{A}^+}(H_0)T_{H_1}^{-1} = \emptyset$$

For finding the double antisymmetry space group types, only conditions for $Q(H)$ and $Q(H)\{K\}$ groups are necessary. This is because it is trivial to map these results to all the other categories of double antisymmetry space groups. This normalizer method is demonstrated in Appendix B to derive all double antisymmetry space group types where $Q = Cc$.

## 6. Number of Types of Double Antisymmetry Space Groups

The total number of types of double antisymmetry space groups listed by the present work is 17,803. The total number of $Q(H)\{K\}$ types this listed by the present work is 9,507. These values differ from those given by Zamorzaev & Palistrant (1980). We have found four fewer **Q(H){K}** types where $Q$ = *Ibca* (serial number 73 on the international tables). Since there are only a small number of discrepancies between our listing and the numbers calculated by Zamorzaev & Palistrant, each will be addressed explicitly.

Zamorzaev & Palistrant (1980) gave a list of double antisymmetry group generators in non-coordinated notation (Koptsik & Shubnikov, 1974). For *Ibca* (*21a* in Zamorzaev & Palistrant (1980))the following generators are used:

$$\left\{a, b, \frac{a+b+c}{2}\right\}\left(\frac{c}{2}2 \cdot \frac{b}{2}m : \frac{a}{2}2_{\frac{b}{4}}\right)$$

The first part, $\left\{a, b, \frac{a+b+c}{2}\right\}$, represents a generating set of three translations for the body-centered lattice of *Ibca*. The second part, $\left(\frac{c}{2}2 \cdot \frac{b}{2}m : \frac{a}{2}2_{\frac{b}{4}}\right)$, generates the remaining operations of *Ibca* when combined with the translations given by the first part.

Zamorzaev & Palistrant (1980) couple these generators with anti-identities to give generating sets for double antisymmetry space groups. However, unlike the more explicit listing given in the present work, Zamorzaev & Palistrant give only generating sets and only those which are unique under the permutations of the elements of **1'1\*** that preserve the group structure, i.e. the automorphisms of **1'1\***. Because of this concise method of listing generating sets (only 1846 **Q(H){K}** generating sets are necessary), a single generating set given by Zamorzaev & Palistrant (Zamorzaev & Palistrant, 1980) can represent up to six types under the proper affine equivalence relation. The six possible types correspond to the six automorphisms of **1'1\***.

The automorphisms of **1'1\*** correspond to the possible permutations of the three anti-identities of double antisymmetry: (1', 1*, 1'*), (1', 1'*, 1*), (1*, 1', 1'*), (1*, 1'*,1'), (1'*, 1', 1*), and (1'*, 1*, 1'), i.e. $\text{Aut}(\mathbf{1'1^*}) \cong \mathbf{S_3}$. Zamorzaev & Palistrant give the number of types represented by each line, but not which automorphisms must be applied to get them. This is discussed in Appendix E.

There are only two lines of generators from Zamorzaev & Palistrant for which their resulting number of **Q(H){K}** types differs from this work. These lines are given in Table 4. In the first line, Zamorzaev & Palistrant give six distinct types whereas there are only three. To show this, we list the six generating sets obtained from the generating set on the first line using the permutations of the anti-identities:

1. $\left\{a, b, \frac{a+b+c}{2}\right\}\left(\frac{c}{2}2'^* \cdot \frac{b}{2}m : \frac{a}{2}2^*_{\frac{b}{4}}\right)$

2. $\left\{a, b, \frac{a+b+c}{2}\right\}\left(\frac{c}{2}2' \cdot \frac{b}{2}m : \frac{a}{2}2^*_{\frac{b}{4}}\right)$

3. $\left\{a, b, \frac{a+b+c}{2}\right\}\left(\frac{c}{2}2'^* \cdot \frac{b}{2}m : \frac{a}{2}2'_{\frac{b}{4}}\right)$

4. $\left\{a, b, \frac{a+b+c}{2}\right\}\left(\frac{c}{2}2^* \cdot \frac{b}{2}m : \frac{a}{2}2'_{\frac{b}{4}}\right)$

5. $\left\{a, b, \frac{a+b+c}{2}\right\}\left(\frac{c}{2}2' \cdot \frac{b}{2}m : \frac{a}{2}2'^*_{\frac{b}{4}}\right)$

6. $\left\{a, b, \frac{a+b+c}{2}\right\}\left(\frac{c}{2}2^* \cdot \frac{b}{2}m : \frac{a}{2}2'^*_{\frac{b}{4}}\right)$

Applying the group-theoretic reasoning shown in Appendix E, for six types, no pair of groups in this set should satisfy the proper affine equivalence relation (represented as ~). To have exactly three types, exactly one of the following must be true:

- 1~2, 3~4, and 5~6
- 1~4, 3~6, and 5~2
- 1~6, 3~2, and 5~4

Recalling that, according to the definition of equivalence, two groups have the same type if they can be mapped into each other by orientation-preserving affine transformations. The present work finds that:

- $(a_2, b_2, c_2) \to \left(a_1 + \frac{1}{4}, c_1 + \frac{1}{4}, -b_1 + \frac{1}{4}\right)$ maps 2 to 1, thus 1~2

- $(a_4, b_4, c_4) \to \left(a_3, c_3 + \frac{1}{2}, -b_3 + \frac{1}{4}\right)$ maps 4 to 3, thus 3~4

- $(a_6, b_6, c_6) \to \left(a_5 + \frac{1}{4}, c_5 + \frac{1}{4}, -b_5 + \frac{1}{4}\right)$ maps 6 to 5, thus 5~6

This satisfies 1~2, 3~4, and 5~6. Thus, contrary to the Zamorzaev & Palistrant listing, there are only three types generated by the first line.

For the second line, Zamorzaev & Palistrant say there are two unique types whereas the present work finds only one. Permutations of the anti-identities give the following six generating sets:

1. $\left\{a, b, \frac{a+b+c}{2}\right\} \left(\frac{c}{2} 2'^* \cdot \frac{b}{2} m^* : \frac{a}{2} 2^*_{\frac{b}{4}}\right)$

2. $\left\{a, b, \frac{a+b+c}{2}\right\} \left(\frac{c}{2} 2' \cdot \frac{b}{2} m^* : \frac{a}{2} 2^*_{\frac{b}{4}}\right)$

3. $\left\{a, b, \frac{a+b+c}{2}\right\} \left(\frac{c}{2} 2'^* \cdot \frac{b}{2} m' : \frac{a}{2} 2'_{\frac{b}{4}}\right)$

4. $\left\{a, b, \frac{a+b+c}{2}\right\} \left(\frac{c}{2} 2^* \cdot \frac{b}{2} m' : \frac{a}{2} 2'_{\frac{b}{4}}\right)$

5. $\left\{a, b, \frac{a+b+c}{2}\right\} \left(\frac{c}{2} 2' \cdot \frac{b}{2} m'^* : \frac{a}{2} 2'^*_{\frac{b}{4}}\right)$

6. $\left\{a, b, \frac{a+b+c}{2}\right\} \left(\frac{c}{2} 2^* \cdot \frac{b}{2} m'^* : \frac{a}{2} 2'^*_{\frac{b}{4}}\right)$

According to the reasoning given in Appendix E, two types implies that 1~4, 1~5, 2~3, 2~6, and 1 ≁ 2. One type implies 1~2, 1~3, 1~4, 1~5, and 1~6. Thus, if Zamorzaev & Palistrant were correct, it should not be possible to map between 1 and 2. Contrary to Zamorzaev & Palistrant, the present work finds that:

- $(a_2, b_2, c_2) \rightarrow \left(a_1 + \frac{1}{4}, c_1 + \frac{1}{4}, -b_1 + \frac{1}{4}\right)$ maps <u>2 to 1</u>

- $(a_3, b_3, c_3) \rightarrow \left(b_1, -a_1 - \frac{1}{2}, c_1 + \frac{1}{4}\right)$ maps <u>3 to 1</u>

- $(a_6, b_6, c_6) \rightarrow \left(c_1 + \frac{1}{4}, b_1 + \frac{1}{4}, -a_1 + \frac{1}{4}\right)$ maps <u>6 to 1</u>

For completeness, the remaining maps (which do not conflict with Zamorzaev & Palistrant's results) follow:

- $(a_4, b_4, c_4) \rightarrow (b_1, c_1, a_1)$ maps 4 to 1

- $(a_5, b_5, c_5) \rightarrow (c_1, a_1, b_1)$ maps 5 to 1

Consequently, the generators in the first line of Table 4 map to three types and the generators in the second line map to one type, not six and two respectively. Thus, there are four fewer **Q(H){K}** types than the number given by Zamorzaev & Palistrant (1980), 9507 rather than 9511. This error likely affects the number of higher order multiple antisymmetry groups calculated by Zamorzaev & Palistrant, as well. We conjecture that there are 24 fewer non-trivial triple antisymmetry space groups then calculated by Zamorzaev & Palistrant but that the numbers for other multiple antisymmetries are correct. If we are correct, this would mean that the number in the final column of Table 1 of "Generalized Antisymmetry" by Zamorzaev (1988) should read 230, 1191, 9507, 109115, 1640955, 28331520, and 419973120, rather than 230, 1191, 9511, 109139, 1640955, 28331520, and 419973120 (the numbers which differ are underlined). Similarly, if we are correct, the final column of Table 2 of the same work should read 230, 1651, 17803, 287574, 6879260, 240768842, and 12209789596 rather than 230, 1651, 17807, 287658, 6880800, 240800462 (typo'ed to be 240900462), and 12210589024.

Our results also confirm that there are 5005 types of **Q(H){K}** Mackay groups (Jablan, 1993a, 2002; Radovic & Jablan, 2005).

### 7. Concluding Remarks

It was found that there are 17,803 types of double antisymmetry groups. This is four fewer than previously stated by Zamorzaev (1988). When rotation-reversal symmetry and time-reversal symmetry are considered, this implies that there are 17,803 distinct types of symmetry a crystal may exhibit.

[†] http://sites.psu.edu/gopalan/research/symmetry/

**Acknowledgements** We acknowledge support from the Penn State Center for Nanoscale Science through the NSF-MRSEC DMR #0820404. We also acknowledge NSF DMR-0908718 and DMR-1210588.

**Table 1**    Double antisymmetry group categories and group structure.

The minus sign, "-", indicates the set-theoretic difference, also known as the relative complement. The plus sign, "+", indicates union. H and K are unique index-2 subgroups of Q (an index-2 subgroup has half as many elements as the group, or equivalently: |Q/H|=2).

| Category | # of types | Class Symbol | | Class Group Structure |
|---|---|---|---|---|
| 1) | 230 | Q | = | Q |
| 2) | 230 | Q1' | = | Q + Q1' |
| 3) | 1191 | Q(H) | = | H + (Q – H)1' |
| 4) | 230 | Q1* | = | Q + Q1* |
| 5) | 230 | Q1'1* | = | Q + Q1' + Q1* + Q1'* |
| 6) | 1191 | Q(H)1* | = | H + (Q-H)1' + H1* + (Q-H)1'* |
| 7) | 1191 | Q{H} | = | H + (Q-H)1* |
| 8) | 230 | Q1'* | = | Q + Q1'* |
| 9) | 1191 | Q{H}1' | = | H + (Q-H)1* + H1' + (Q-H)1'* |
| 10) | 1191 | Q(H)1'* | = | H + (Q-H)1' + H1'* + (Q-H)1* |
| 11) | 1191 | Q(H){H} | = | H + (Q-H)1'* |
| 12) | 9507 | Q(H){K} | = | H∩K + (H-K)1* + (K-H)1' + (Q-(H+K))1'* |
| Total | 17,803 | | | |

**Table 2** Examples of listings in Seitz notation.

a)

No. 8543   C2'/m*
(1 | 0 0 0)  (2$_y$ | 0 0 0)'  (m$_y$ | 0 0 0)*  ($\bar{1}$ | 0 0 0)'*

b)

No. 16490   I4*/mm'm'*
(1 | 0 0 0)  ($\bar{1}$ | 0 0 0)  (2$_z$ | 0 0 0)  (m$_z$ | 0 0 0)  (2$_x$ | 0 0 0)'  (2$_y$ | 0 0 0)'
(m$_x$ | 0 0 0)'  (m$_y$ | 0 0 0)'  (4$_z$ | 0 0 0)*  ($\bar{4}_z$ | 0 0 0)*  (4$_z^{-1}$ | 0 0 0)*  ($\bar{4}_z^{-1}$ | 0 0 0)*
(2$_{x\bar{y}}$ | 0 0 0)'*  (2$_{xy}$ | 0 0 0)'*  (m$_{x\bar{y}}$ | 0 0 0)'*  (m$_{xy}$ | 0 0 0)'*

c)

No. 13461    Ib'*c'a'

(1 | 0 0 0)   (2$_x$ | 0 0 1/2)   (m$_y$ | 1/2 0 0)'   (m$_z$ | 1/2 0 1/2)'   (2$_y$ | 1/2 0 0)*   (2$_z$ | 1/2 0 1/2)*
($\bar{1}$ | 0 0 0)'*   (m$_x$ | 0 0 1/2)'*

**Table 3**   Coloring of lattice symbols

| Lattice symbol **P(1',1*,1'*)** | First position P(**1'**,1*,1'*) | Second position P(1',**1***,1'*) | Third position P(1',1*,**1'***) |
|---|---|---|---|
| P | t$_{[100]}$ | t$_{[010]}$ | t$_{[001]}$ |
| C | t$_{[100]}$ | t$_{[001]}$ | t$_{[½½0]}$ |
| A | t$_{[100]}$ | t$_{[010]}$ | t$_{[0½½]}$ |
| I | t$_{[100]}$ | t$_{[001]}$ | t$_{[½½½]}$ |
| F | t$_{[0½½]}$ | t$_{[½0½]}$ | t$_{[½½0]}$ |
| R | t$_{[001]}$ | t$_{[⅔⅓⅓]}$ | t$_{[⅓⅔⅔]}$ |

**Table 4**   Double antisymmetry space group generating sets listed by Zamorzaev & Palistrant (1980).

The numbers in the second and third columns refer to the number of unique types obtained by the permutation of anti-identities.

| Generating line | Zamorzaev & Palistrant (1980) | Actual (present work) | Difference |
|---|---|---|---|
| $\left\{a, b, \dfrac{a+b+c}{2}\right\}\left(\dfrac{c}{2}2' \cdot \dfrac{b}{2}m : \dfrac{a}{2}2^*_{\frac{b}{4}}\right)$ | 6 | 3 | -3 |
| $\left\{a, b, \dfrac{a+b+c}{2}\right\}\left(\dfrac{c}{2}2' \cdot \dfrac{b}{2}m^* : \dfrac{a}{2}2^*_{\frac{b}{4}}\right)$ | 2 | 1 | -1 |

**Figure 1**  Identity (1) and anti-identities (1', 1*, and 1'*) of the rotation-reversal and time-reversal space groups.

**Figure 2**  a) Multiplication table of **1'1***. To evaluate the product of two elements, we find the row associated with the first element and the column associated with second element, e.g. for 1'·1*, go to the second row third column to find 1'*. b) Cayley graph generated by 1', 1*, and 1'*. To evaluate the product of two elements, we start from the circle representing the first element and follow the arrow representing the second, e.g. for 1'·1*, we start on the red circle (1') and take the blue path (1*) to the green circle (1'*).

**Figure 3** Demonstration of proper affine equivalence of **Q(H){K}** groups generated for *Q* = **222** using point group diagrams (stereoscopic projections).

**Figure 4** Example double antisymmetry space group diagrams. **a)** No. 8543 C2'/m*, **b)** No. 16490 I4*/mm'm'*, **c)** No. 13461, Ib'*c'a'

**Appendix A. Visual representation of category structures using Venn diagrams**

We can visually represent the set structure of each category, starting by defining Q as 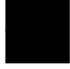, H as 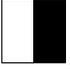, and K as 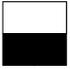. It follows that set-theoretic difference Q-H is 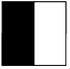. Likewise, H∩K, H-K, K-H, and Q-(H+K) are 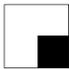, 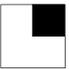, 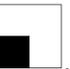, and 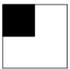 respectively. When the elements of Q are colored with an anti-identity, the resulting set will be represented by a square of the corresponding color, i.e. Q1' = 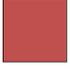, Q1*= 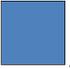, and Q1'*= 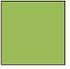. Following this scheme, the twelve categories of double antisymmetry groups can be represented as follows:

| Category | Category Symbol | | Category Structure |
|---|---|---|---|
| 1) | Q | = | Q |
| | 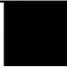 | = | 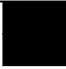 |
| 2) | Q1' | = | Q + Q1' |
| | 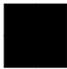 + 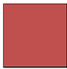 | = | 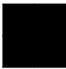 + 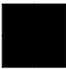 1' |
| 3) | Q(H) | = | H + (Q – H)1' |
| | 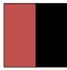 | = | 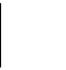 + 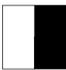 1' |
| 4) | Q1* | = | Q + Q1* |
| | 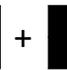 + 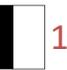 | = | 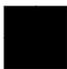 + 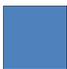 1* |
| 5) | Q1'1* | = | Q + Q1' + Q1* + Q1'* |
| | 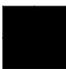 + 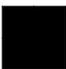 + 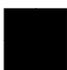 + 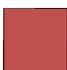 | = | 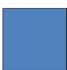 + 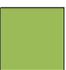 1' + 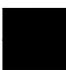 1* + 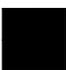 1'* |
| 6) | Q(H)1* | = | H + (Q-H)1' + H1* + (Q-H)1'* |
| | 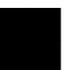 + 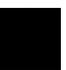 | = | 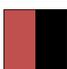 + 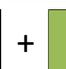 1' + 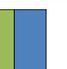 1* + 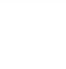 1'* |
| 7) | Q{H} | = | H + (Q-H)1* |
| | 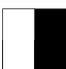 | = | 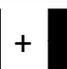 + 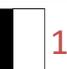 1* |
| 8) | Q1'* | = | Q + Q1'* |
| | 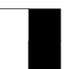 + 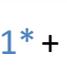 | = | 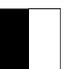 + 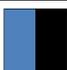 1'* |
| 9) | Q{H}1' | = | H + (Q-H)1* + H1' + (Q-H)1'* |
| | 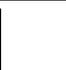 + 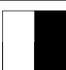 | = | 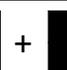 + 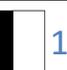 1*+ 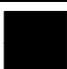 1' + 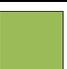 1'* |
| 10) | Q(H)1'* | = | H + (Q-H)1' + H1'* + (Q-H)1* |
| | 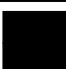 + 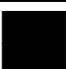 | = | 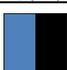 + 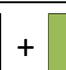 1' + 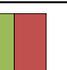 1'*+ 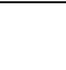 1* |
| 11) | Q(H){H} | = | H + (Q-H)1'* |
| | 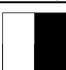 | = | 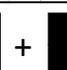 + 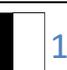 1'* |
| 12) | Q(H){K} | = | H∩K + (H-K)1* + (K-H)1' + (Q-(H+K))1'* |
| | 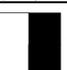 | = | 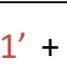 + 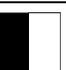 1*+ 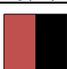 1' + 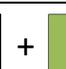 1'* |

## Appendix B. Additional examples of generating double antisymmetry groups

**2/m** has four elements: {1,2,m,-1}. **2/m** has three index-2 subgroups: {1,2}, {1,m}, and {1,-1} which will be referred to as **2**, **m**, and **-1** respectively.

Category 1: **Q**

**Q** = 2/m → **Q** = {1,2,m,-1}

Category 2: **Q** + **Q**1'

**Q** = 2/m → **Q**1' = {1,2,m,-1,1',2',m',-1'}

Category 3: **H** + (**Q** – **H**)1'

**Q** = 2/m, **H** = 2 → **Q**(**H**) = {1,2,m',-1'}

**Q** = 2/m, **H** = m → **Q**(**H**) = {1,2',m,-1'}

**Q** = 2/m, **H** = -1 → **Q**(**H**) = {1,2',m',-1}

Category 4: **Q** + **Q**1*

**Q** = 2/m → **Q**1* = {1,2,m,-1,1*,2*,m*,-1*}

Category 5: **Q** + **Q**1' + **Q**1* + **Q**1'*

**Q** = 2/m → **Q**1'1* = {1,2,m,-1,1',2',m',-1',1*,2*,m*,-1*,1'*,2'*,m'*,-1'*}

Category 6: **H** + (**Q**-**H**)1' + **H**1* + (**Q**-**H**)1'*

**Q** = 2/m, **H** = 2 → **Q**(**H**)1* = {1,2,m',-1',1*,2*,m'*,-1'*}

**Q** = 2/m, **H** = m → **Q**(**H**)1* = {1,2',m,-1',1*,2'*,m*,-1'*}

**Q** = 2/m, **H** = -1 → **Q**(**H**)1* = {1,2',m',-1,1*,2'*,m'*,-1*}

Category 7: **H** + (**Q**-**H**)1*

**Q** = 2/m, **H** = 2 → **Q**{**H**} = {1,2,m*,-1*}

**Q** = 2/m, **H** = m → **Q**{**H**} = {1,2*,m,-1*}

**Q** = 2/m, **H** = -1 → **Q**{**H**} = {1,2*,m*,-1}

Category 8: **Q** + **Q**1'*

**Q** = 2/m → **Q**1'* = {1,2,m,-1,1'*,2'*,m'*,-1'*}

Category 9: **H** + (**Q**-**H**)1* + **H**1' + (**Q**-**H**)1'*

$Q$ = 2/m, $H$ = 2 → $Q\{H\}1'$ = {1,2,m*,-1*,1',2',m'*,-1'*}

$Q$ = 2/m, $H$ = m → $Q\{H\}1'$ = {1,2*,m,-1*,1',2'*,m',-1'*}

$Q$ = 2/m, $H$ = -1 → $Q\{H\}1'$ = {1,2*,m*,-1,1',2'*,m'*,-1'}

Category 10: $H + (Q-H)1' + H1'^* + (Q-H)1^*$

$Q$ = 2/m, $H$ = 2 → $Q(H)1'^*$ = {1,2,m',-1',1'*,2'*,m*,-1*}

$Q$ = 2/m, $H$ = m → $Q(H)1'^*$ = {1,2',m,-1',1'*,2*,m'*,-1*}

$Q$ = 2/m, $H$ = -1 → $Q(H)1'^*$ = {1,2',m',-1,1'*,2*,m*,-1'*}

Category 11: $H + (Q-H)1^*$

$Q$ = 2/m, $H$ = 2 → $Q(H)\{H\}$ = {1,2,m'*,-1'*}

$Q$ = 2/m, $H$ = m → $Q(H)\{H\}$ = {1,2'*,m,-1'*}

$Q$ = 2/m, $H$ = -1 → $Q(H)\{H\}$ = {1,2'*,m'*,-1}

Category 12: $H \cap K + (H-K)1^* + (K-H)1' + (Q-(H+K))1'^*$

$Q$ = 2/m, $H$ = m, $K$ = -1 → $Q(H)\{K\}$ = {1,2'*,m*,-1'}

$Q$ = 2/m, $H$ = 2, $K$ = -1 → $Q(H)\{K\}$ = {1,2*,m'*,-1'}

$Q$ = 2/m, $H$ = 2, $K$ = m → $Q(H)\{K\}$ = {1,2*,m',-1'*}

$Q$ = 2/m, $H$ = -1, $K$ = m → $Q(H)\{K\}$ = {1,2'*,m',-1*}

$Q$ = 2/m, $H$ = -1, $K$ = 2 → $Q(H)\{K\}$ = {1,2',m'*,-1*}

$Q$ = 2/m, $H$ = m, $K$ = 2 → $Q(H)\{K\}$ = {1,2',m*,-1'*}

Both **2/m** and **222** (given as an example in Section 2) have three index-2 subgroups. Consequently, they generate the same number of double antisymmetry groups: 29. However, unlike with **222**, none of the 29 groups formed from **2/m** are in the same equivalence class. This may seem surprising given that **2/m** is isomorphic to **222**. This can be thought of as being a consequence of the fact that none of the elements of **2/m** can be rotated into one another whereas the three two-fold axes of **222** can be. Another way to look at it is to consider that **2/m**'s proper affine normalizer group (∞2) does not contain non-trivial automorphisms of whereas **222**'s proper affine normalizer group (**432**) does.

As with all crystallographic space groups, **Cc** has an infinite number of elements due to the infinite translational subgroup. **Cc**'s elements will be represented as $(t_{[100]}, t_{[001]}, t_{[½½0]})\{1,c\}$ where $(t_{[100]}, t_{[001]}, t_{[½½0]})$ represent the generators of the translation subgroup and $\{1,c\}$ are the coset representatives of corresponding decomposition. **Cc** has three index-2 subgroups: $(t_{[100]},t_{[001]},t_{[½½0]})\{1\}$, $(t_{[100]},t_{[010]},t_{[001]})\{1,c\}$, and $(t_{[100]},t_{[010]},t_{[001]})\{1, t_{[½½0]} \cdot c\}$.

Category 1) **Q**

**Q** = **Cc** → **Q** = $(t_{[100]},t_{[001]},t_{[½½0]})\{1,c\}$

Category 2) **Q** + **Q**1'

**Q** = **Cc** → **Q1'** = $(t_{[100]},t_{[001]},t_{[½½0]})\{1,c,1',c'\}$

Category 3) **H** + (**Q** – **H**)1'

**Q** = **Cc**, **H** = $(t_{[100]},t_{[001]},t_{[½½0]})\{1\}$ → **Q(H)** = $(t_{[100]},t_{[001]},t_{[½½0]})\{1,c'\}$

**Q** = **Cc**, **H** = $(t_{[100]},t_{[010]},t_{[001]})\{1,c\}$ → **Q(H)** = $(t_{[100]},t_{[001]},t_{[½½0]}')\{1,c\}$

**Q** = **Cc**, **H** = $(t_{[100]},t_{[010]},t_{[001]})\{1,t_{[½½0]} \cdot c\}$ → **Q(H)** = $(t_{[100]},t_{[001]},t_{[½½0]}')\{1,c'\}$

Category 4) **Q** + **Q**1*

**Q** = **Cc** → **Q1*** = $(t_{[100]},t_{[001]},t_{[½½0]})\{1,c,1*,c*\}$

Category 5) **Q** + **Q**1' + **Q**1* + **Q**1'*

**Q** = **Cc** → **Q1'1*** = $(t_{[100]},t_{[001]},t_{[½½0]})\{1,c,1',c',1*,c*,1'*,c'*\}$

Category 6) **H** + (**Q-H**)1' + **H**1* + (**Q-H**)1'*

**Q** = **Cc**, **H** = $(t_{[100]},t_{[001]},t_{[½½0]})\{1\}$ → **Q(H)1*** = $(t_{[100]},t_{[001]},t_{[½½0]})\{1,c',1*,c'*\}$

**Q** = **Cc**, **H** = $(t_{[100]},t_{[010]},t_{[001]})\{1,c\}$ → **Q(H)1*** = $(t_{[100]},t_{[001]},t_{[½½0]}')\{1,c,1*,c*\}$

**Q** = **Cc**, **H** = $(t_{[100]},t_{[010]},t_{[001]})\{1,t_{[½½0]} \cdot c\}$ → **Q(H)1*** = $(t_{[100]},t_{[001]},t_{[½½0]}')\{1,c',1*,c'*\}$

Category 7) **H** + (**Q-H**)1*

**Q** = **Cc**, **H** = $(t_{[100]},t_{[001]},t_{[½½0]})\{1\}$ → **Q{H}** = $(t_{[100]},t_{[001]},t_{[½½0]})\{1,c*\}$

**Q** = **Cc**, **H** = $(t_{[100]},t_{[010]},t_{[001]})\{1,c\}$ → **Q{H}** = $(t_{[100]},t_{[001]},t_{[½½0]}*)\{1,c\}$

**Q** = **Cc**, **H** = $(t_{[100]},t_{[010]},t_{[001]})\{1,t_{[½½0]} \cdot c\}$ → **Q{H}** = $(t_{[100]},t_{[001]},t_{[½½0]}*)\{1,c*\}$

Category 8) **Q** + **Q**1'*

**Q** = **Cc** → **Q1'*** = $(t_{[100]},t_{[001]},t_{[½½0]})\{1,c,1'*,c'*\}$

Category 9) **H** + (**Q-H**)1* + **H**1' + (**Q-H**)1'*

**Q** = **Cc**, **H** = $(t_{[100]},t_{[001]},t_{[½½0]})\{1\}$ → **Q{H}1*** = $(t_{[100]},t_{[001]},t_{[½½0]})\{1,c*,1',c'*\}$

$Q = \mathbf{Cc}$, $H = (t_{[100]},t_{[010]},t_{[001]})\{1,c\} \rightarrow Q\{H\}1^* = (t_{[100]},t_{[001]},t_{[½½0]}{}^*)\{1,c,1',c'\}$

$Q = \mathbf{Cc}$, $H = (t_{[100]},t_{[010]},t_{[001]})\{1,t_{[½½0]}\cdot c\} \rightarrow Q\{H\}1^* = (t_{[100]},t_{[001]},t_{[½½0]}{}^*)\{1,c^*,1',c'^*\}$

Category 10) $H + (Q-H)1' + H1'^* + (Q-H)1^*$

$Q = \mathbf{Cc}$, $H = (t_{[100]},t_{[001]},t_{[½½0]})\{1\} \rightarrow Q(H)1'^* = (t_{[100]},t_{[001]},t_{[½½0]})\{1,c',1'^*,c^*\}$

$Q = \mathbf{Cc}$, $H = (t_{[100]},t_{[010]},t_{[001]})\{1,c\} \rightarrow Q(H)1'^* = (t_{[100]},t_{[001]},t_{[½½0]}')\{1,c,1'^*,c'^*\}$

$Q = \mathbf{Cc}$, $H = (t_{[100]},t_{[010]},t_{[001]})\{1,t_{[½½0]}\cdot c\} \rightarrow Q(H)1'^* = (t_{[100]},t_{[001]},t_{[½½0]}')\{1,c',1'^*,c^*\}$

Category 11) $H + (Q-H)1'^*$

$Q = \mathbf{Cc}$, $H = (t_{[100]},t_{[001]},t_{[½½0]})\{1\} \rightarrow Q\{H\} = (t_{[100]},t_{[001]},t_{[½½0]})\{1,c'^*\}$

$Q = \mathbf{Cc}$, $H = (t_{[100]},t_{[010]},t_{[001]})\{1,c\} \rightarrow Q\{H\} = (t_{[100]},t_{[001]},t_{[½½0]}{}^*)\{1,c\}$

$Q = \mathbf{Cc}$, $H = (t_{[100]},t_{[010]},t_{[001]})\{1,t_{[½½0]}\cdot c\} \rightarrow Q\{H\} = (t_{[100]},t_{[001]},t_{[½½0]}{}^*)\{1,c^*\}$

Category 12) $H \cap K + (H-K)1^* + (K-H)1' + (Q-(H+K))1'^*$

$Q = \mathbf{Cc}$, $H = (t_{[100]},t_{[010]},t_{[001]})\{1,c\}$, $K = (t_{[100]},t_{[010]},t_{[001]})\{1,t_{[½½0]}\cdot c\} \rightarrow Q(H)\{K\} = (t_{[100]},t_{[001]},t_{[½½0]}{}^*)\{1,c^*\}$

$Q = \mathbf{Cc}$, $H = (t_{[100]},t_{[001]},t_{[½½0]})\{1\}$, $K = (t_{[100]},t_{[010]},t_{[001]})\{1,t_{[½½0]}\cdot c\} \rightarrow Q(H)\{K\} = (t_{[100]},t_{[001]},t_{[½½0]}{}^*)\{1,c'^*\}$

$Q = \mathbf{Cc}$, $H = (t_{[100]},t_{[001]},t_{[½½0]})\{1\}$, $K = (t_{[100]},t_{[010]},t_{[001]})\{1,c\} \rightarrow Q(H)\{K\} = (t_{[100]},t_{[001]},t_{[½½0]}{}^*)\{1,c'\}$

$Q = \mathbf{Cc}$, $H = (t_{[100]},t_{[010]},t_{[001]})\{1,t_{[½½0]}\cdot c\}$, $K = (t_{[100]},t_{[010]},t_{[001]})\{1,c\} \rightarrow Q(H)\{K\} = (t_{[100]},t_{[001]},t_{[½½0]}{}^*)\{1,c'\}$

$Q = \mathbf{Cc}$, $H = (t_{[100]},t_{[010]},t_{[001]})\{1,t_{[½½0]}\cdot c\}$, $K = (t_{[100]},t_{[001]},t_{[½½0]})\{1\} \rightarrow Q(H)\{K\} = (t_{[100]},t_{[001]},t_{[½½0]}')\{1,c'^*\}$

$Q = \mathbf{Cc}$, $H = (t_{[100]},t_{[010]},t_{[001]})\{1,c\}$, $K = (t_{[100]},t_{[001]},t_{[½½0]})\{1\} \rightarrow Q(H)\{K\} = (t_{[100]},t_{[001]},t_{[½½0]}')\{1,c^*\}$

Note that although 29 double antisymmetry space groups are generated from using **Cc** as a colorblind parent group, they are not all of unique types. This is because there exist proper affine normalizers of **Cc** which map some of these into each other. We show this by applying the results of section 5.3.

To do this, we need to know the transformation matrices ($T_{H_1}$ and $T_{H_2}$) mapping the standard representation of the subgroup types ($H_0$) to the actual subgroups ($H_1$ and $H_2$), and the proper affine normalizer groups of $Q$ and $H_0$. For **Cc**'s three index-2 subgroups:

$(t_{[100]},t_{[001]},t_{[½½0]})\{1\}$ is type **P1** and can be mapped from the standard **P1** by $\begin{pmatrix} 1/2 & 1/2 & 0 & 0 \\ -1/2 & 1/2 & 0 & 0 \\ 0 & 0 & 1 & 0 \\ 0 & 0 & 0 & 1 \end{pmatrix}$

$(t_{[100]}, t_{[010]}, t_{[001]})\{1,c\}$ is type **Pc** and can be mapped from the standard **Pc** by $\begin{pmatrix} 1 & 0 & 0 & 0 \\ 0 & 1 & 0 & 0 \\ 0 & 0 & 1 & 0 \\ 0 & 0 & 0 & 1 \end{pmatrix}$

$(t_{[100]}, t_{[010]}, t_{[001]})\{1, t_{[\frac{1}{2}\frac{1}{2}0]} \cdot c\}$ is type **Pc** and can be mapped from the standard **Pc** by $\begin{pmatrix} 1 & 0 & -1 & 0 \\ 0 & 1 & 0 & 1/4 \\ 0 & 0 & 1 & 0 \\ 0 & 0 & 0 & 1 \end{pmatrix}$

The proper affine normalizers of **Cc** are:

$$N_{\mathcal{A}^+}(Cc) = \left\{ \begin{pmatrix} 2n_1 + 1 & 0 & 2n_2 + p & r \\ 0 & \pm 1 & 0 & (2n_3 + p)/4 \\ 2n_4 & 0 & 2n_5 + 1 & t \\ 0 & 0 & 0 & 1 \end{pmatrix} \in \mathcal{A}^+ : r, t \in \mathbb{R} \land n_i \in \mathbb{Z} \land (p = 0 \lor p = 1) \right\}$$

The proper affine normalizers of the standard representations of the two types of subgroups (**P1** and **Pc**) are:

**P1** normalizers: $N_{\mathcal{A}^+}(P1) = \left\{ \begin{pmatrix} n_{11} & n_{12} & n_{13} & r \\ n_{21} & n_{22} & n_{23} & s \\ n_{31} & n_{32} & n_{33} & t \\ 0 & 0 & 0 & 1 \end{pmatrix} \in \mathcal{A}^+ : r, s, t \in \mathbb{R} \land n_{ij} \in \mathbb{Z} \right\}$

**Pc** normalizers: $N_{\mathcal{A}^+}(Pc) = \left\{ \begin{pmatrix} 2n_6 + 1 & 0 & 2n_7 & r \\ 0 & \pm 1 & 0 & n_8/2 \\ n_9 & 0 & 2n_{10} + 1 & t \\ 0 & 0 & 0 & 1 \end{pmatrix} \in \mathcal{A}^+ : r, t \in \mathbb{R} \land n_i \in \mathbb{Z} \right\}$

Having collected all this information, we can now evaluate the proper affine equivalence of the 29 double antisymmetry groups generated from $Q = Cc$.

We know that groups from different categories can never be equivalent; therefore categories 1), 2), 4), 5) and 8) must contain only one type as only one group has been generated.

For category 3), we have three generated groups. Thus there are three pairs for which we can test for equivalence, 3~4, 3~5, and 4~5. For 3, $H$ is **P1** type whereas for 4 and 5 $H$ is **Pc** type. Therefore, 3~4 and 3~5 are false. For 4~5, we can evaluate:

**Equation 8**

$$Q(H_2) \sim Q(H_1) \equiv N_{\mathcal{A}^+}(Q) \cap T_{H_2} N_{\mathcal{A}^+}(H_0) T_{H_1}^{-1} = \emptyset$$

In this case,

$$T_{H_2} N_{\mathcal{A}^+}(H_0) T_{H_1}^{-1} = \left\{ \begin{pmatrix} 1 & 0 & -1 & 0 \\ 0 & 1 & 0 & \frac{1}{4} \\ 0 & 0 & 1 & 0 \\ 0 & 0 & 0 & 1 \end{pmatrix} \begin{pmatrix} 2n_6 + 1 & 0 & 2n_7 & r \\ 0 & \pm 1 & 0 & n_8/2 \\ n_9 & 0 & 2n_{10} + 1 & t \\ 0 & 0 & 0 & 1 \end{pmatrix} \begin{pmatrix} 1 & 0 & 0 & 0 \\ 0 & 1 & 0 & 0 \\ 0 & 0 & 1 & 0 \\ 0 & 0 & 0 & 1 \end{pmatrix}^{-1} \in \mathcal{A}^+ : r, t \in \mathbb{R} \land n_i \in \mathbb{Z} \right\}$$

and

$$N_{\mathcal{A}^+}(Q) = \left\{ \begin{pmatrix} 2n_1+1 & 0 & 2n_2+p & r \\ 0 & \pm 1 & 0 & (2n_3+p)/4 \\ 2n_4 & 0 & 2n_5+1 & t \\ 0 & 0 & 0 & 1 \end{pmatrix} \in \mathcal{A}^+ : r,t \in \mathbb{R} \land n_i \in \mathbb{Z} \land (p=0 \lor p=1) \right\}$$

From substituting these in and simplifying, we can show that 4~5 is logically equivalent to the existence of a solution to the following:

**Equation 9**

$$\begin{pmatrix} 2n_1+1 & 0 & 2n_2+p & r \\ 0 & \pm 1 & 0 & (2n_3+p)/4 \\ 2n_4 & 0 & 2n_5+1 & t \\ 0 & 0 & 0 & 1 \end{pmatrix} = \begin{pmatrix} 2n_6+1-n_9 & 0 & -1-2n_{10}+2n_7 & r-t \\ 0 & \pm 1 & 0 & (2n_8+1)/4 \\ n_9 & 0 & 2n_{10}+1 & t \\ 0 & 0 & 0 & 1 \end{pmatrix}$$

There are clearly many solutions, e.g. one solution is where $n_2 = -1$, $n_{i \neq 2} = r = t = 0$, and $p = 1$. Thus, 4 and 5 are equivalent and therefore for category 3), there are only two types of groups where $Q = Cc$. It is trivial to extend these results to show that categories 6), 7), 9), 10), and 11) similarly have two types.

For category 12), we have six generated groups. Thus there are fifteen pairs for which we can test for equivalence. Only three of the fifteen have the same $H$ and $K$ types (26~25, 24~27, and 28~29) and therefore only these need to be evaluated using:

**Equation 10**

$$Q(H_2)\{K_2\} \sim Q(H_1)\{K_1\} \equiv N_{\mathcal{A}^+}(Q) \cap T_{H_2} N_{\mathcal{A}^+}(H_0) T_{H_1}^{-1} \cap T_{K_2} N_{\mathcal{A}^+}(K_0) T_{K_1}^{-1} \neq \emptyset$$

For 26~25,

$$T_{H_2} N_{\mathcal{A}^+}(H_0) T_{H_1}^{-1} = \left\{ \begin{pmatrix} 1 & 0 & -1 & 0 \\ 0 & 1 & 0 & \frac{1}{4} \\ 0 & 0 & 1 & 0 \\ 0 & 0 & 0 & 1 \end{pmatrix} \begin{pmatrix} 2n_6+1 & 0 & 2n_7 & r \\ 0 & \pm 1 & 0 & n_8/2 \\ n_9 & 0 & 2n_{10}+1 & t \\ 0 & 0 & 0 & 1 \end{pmatrix} \begin{pmatrix} 1 & 0 & 0 & 0 \\ 0 & 1 & 0 & 0 \\ 0 & 0 & 1 & 0 \\ 0 & 0 & 0 & 1 \end{pmatrix}^{-1} \in \mathcal{A}^+ : r,t \in \mathbb{R} \land n_i \in \mathbb{Z} \right\},$$

$$T_{K_2} N_{\mathcal{A}^+}(K_0) T_{K_1}^{-1} = \left\{ \begin{pmatrix} \frac{1}{2} & \frac{1}{2} & 0 & 0 \\ -\frac{1}{2} & \frac{1}{2} & 0 & 0 \\ 0 & 0 & 1 & 0 \\ 0 & 0 & 0 & 1 \end{pmatrix} \begin{pmatrix} n_{11} & n_{12} & n_{13} & r \\ n_{21} & n_{22} & n_{23} & s \\ n_{31} & n_{32} & n_{33} & t \\ 0 & 0 & 0 & 1 \end{pmatrix} \begin{pmatrix} \frac{1}{2} & \frac{1}{2} & 0 & 0 \\ -\frac{1}{2} & \frac{1}{2} & 0 & 0 \\ 0 & 0 & 1 & 0 \\ 0 & 0 & 0 & 1 \end{pmatrix}^{-1} \in \mathcal{A}^+ : r,s,t \in \mathbb{R} \land n_{ij} \in \mathbb{Z} \right\},$$

and

$$N_{\mathcal{A}^+}(Q) = \left\{ \begin{pmatrix} 2n_1+1 & 0 & 2n_2+p & r \\ 0 & \pm 1 & 0 & (2n_3+p)/4 \\ 2n_4 & 0 & 2n_5+1 & t \\ 0 & 0 & 0 & 1 \end{pmatrix} \in \mathcal{A}^+ : r,t \in \mathbb{R} \land n_i \in \mathbb{Z} \land (p=0 \lor p=1) \right\}.$$

From substituting these in and simplifying, we can show that 26~25 is logically equivalent to the existence of a solution to the following:

Equation 11

$$\begin{pmatrix} 2n_1+1 & 0 & 2n_2+p & r \\ 0 & \pm1 & 0 & (2n_3+p)/4 \\ 2n_4 & 0 & 2n_5+1 & t \\ 0 & 0 & 0 & 1 \end{pmatrix} = \begin{pmatrix} 2n_6+1-n_9 & 0 & -1-2n_{10}+2n_7 & r-t \\ 0 & \pm1 & 0 & (2n_8+1)/4 \\ n_9 & 0 & 2n_{10}+1 & t \\ 0 & 0 & 0 & 1 \end{pmatrix} =$$

$$\begin{pmatrix} n_{11}-n_{31} & n_{12}-n_{32} & n_{13}-n_{33} & r-t \\ n_{21} & n_{22} & n_{23} & s+1/4 \\ n_{31} & n_{32} & n_{33} & t \\ 0 & 0 & 0 & 1 \end{pmatrix}$$

There are clearly many solutions, e.g. one solution is where $n_2 = -1$, $n_{i \neq 2} = r = t = s = 0$, $p = 1$, $n_{i=j} = 1$, and $n_{i \neq j} = 0$. Thus 25 and 26 are equivalent. Since 24, 27, 28 and 29 can be related to 25 and 26 by automorphisms of **1'1***, 26~25 implies 24~27 and 28~29. Therefore there are only three types of category 12) groups and a total of twenty double antisymmetry group types for $Q = Cc$.

## Appendix C. Equivalence classes, proper affine classes (types), Mackay groups, and color-permuting classes

An *equivalence relation* can be used to partition a set of groups into *equivalence classes*. For example, an equivalence relation can be applied to partition the set of crystallographic space groups (which is uncountably infinite) into a finite number of classes. The proper affine equivalence relation is used to classify space groups into 230 *proper affine classes* or "*types*". Proper affine equivalence relation be defined as: two groups, $G_1$ and $G_2$, are equivalent if and only if $G_1$ can be bijectively mapped to $G_2$ by a proper affine transformation, $a$.

Equation 12

$$G_2 \sim G_1 \equiv \exists a \in \mathcal{A}^+ : (G_2 = a G_1 a^{-1})$$

If it is known that $G_1$ and $G_2$ have the same colorblind parent group $Q$, then, instead of using the entire proper affine group $\mathcal{A}^+$, it is sufficient to use the proper affine normalizer group of $Q$, denoted $N_{\mathcal{A}^+}(Q)$.

The proper affine equivalence relation does not allow for any permutations of anti-identities. Other works give another set of equivalence classes of antisymmetry groups called Mackay groups (S. V. Jablan, 1993a, 2002; Radovic & Jablan, 2005). The equivalence relation of Mackay groups allows some color permutations in addition to proper affine transformation. For double antisymmetry, the Mackay equivalence relation allows for 1' and 1* to be permutated, i.e. all the primed operations become starred and vice versa:

**Equation 13**

$$G_2 \sim G_1 \equiv \exists a \in \mathcal{A}^+ \wedge \exists p \in \{1, 1' \leftrightarrow 1^*\} : (G_2 = ap(G_1)a^{-1})$$

The Mackay equivalence relation does not allow for 1'* to be permutated (Radovic & Jablan, 2005). Note that Radovic & Jablan do give the Mackay equivalence relation as permuting "anti-identities" but 1'* is not considered an anti-identity in their work (it is simply the product of 1' and 1*). They also conclude that Mackay groups are the minimal representation of "Zamorzaev groups". This seems potentially inconsistent with the aforementioned restriction on color permutation. If we instead allow for all possible color permutations that preserve the group structure of **1'1***, i.e. the automorphisms of **1'1***, we can clearly further reduce the representation beyond that of the Mackay groups, contrary to what has been claimed. This is demonstrated by Zamorzaev & Palistrant's listing of double antisymmetry space group generating sets. In their listing, they gave only those set which were unique up to the automorphisms of **1'1*** (Zamorzaev & Palistrant, 1980). Such a listing only needs to contain 1846 **Q(H){K}** generating sets, far fewer than the 5005 Mackay equivalence classes of **Q(H){K}** groups.

If all possible color permutations that preserve the group structure of **1'1***, i.e. the automorphisms of **1'1***, are allowed the equivalence relation is expressed as:

**Equation 14**

$$G_2 \sim G_1 \equiv \exists a \in \mathcal{A}^+ \wedge \exists p \in \mathrm{Aut}(\mathbf{1'1^*}) : (G_2 = ap(G_1)a^{-1})$$

This proper affine *color equivalence relation* results in 1846 classes for category 12) **Q(H){K}** groups.

Generalized to an arbitrary coloring scheme, *P*, the *color equivalence relation* be defined as:

**Equation 15**

$$G_2 \sim G_1 \equiv \exists a \in \mathcal{A}^+ \times \mathrm{Aut}(P) : (G_2 = aG_1a^{-1})$$

The advantage of using the color equivalence relation to reduce the number of equivalence classes becomes greater as the number of colors (the order of *P*) increases. For example, for non-trivial sextuple antisymmetry space groups (i.e. $P \simeq \mathbb{Z}_2^6$), there are 419,973,120 proper affine equivalence classes and 598,752 Mackay equivalence classes, but just 1 color equivalence class.

Although these color permuting equivalence relations reduce the number of equivalence classes significantly, they are not suitable when the differences between the anti-identities are important. With time-reversal as 1' and rotation-reversal as 1*, the differences are clearly very important. However, there may be applications where the color equivalence relation is suitable, for example, in making patterns for aesthetic purposes.

## Appendix D. Using the machine-readable file

All 38,290 double antisymmetry space groups generated by applying formulae in Table 1 to the standard representative of each of the 230 crystallographic space group types are listed in "DASGMachineReadable.txt". Each group is represented as eight lines; the entire file contains a total of 306,320 lines (8*38,290).

The first line of each set of eight contains three numbers: the serial number of the type (1 to 17,803), the "setting" number to identify groups with the same type (1 to the number of groups in that type), and the number of the category of the group (1 to 12).

The second line contains five numbers: the space group number of $Q$ (1 to 230), the space group number of $H$ (1 to 230), the space group number of $K$ (1 to 230), the space group number of $L$ (1 to 230), and the space group number of $R$ (1 to 230). $Q$, $H$, and $K$, have the same meaning as previously given in Table 1. $R$ is $H \cap K$ and is therefore an index-4 subgroup of $Q$. $L$ is an index-2 subgroup of $Q$ which is equivalent to $Q-(H+K)+R$. If a number is not applicable to the current category then "0" is given.

The third line contains 80 numbers. These are to be partitioned into five 4-by-4 matrices representing the transformations from $Q_0$, $H_0$, $K_0$, $L_0$, and $R_0$, onto $Q$, $H$, $K$, $L$, and $R$ respectively, e.g. the third matrix is $T_H$ such that $H = T_H H_0 T_H^{-1}$. $Q_0$, $H_0$, $K_0$, $L_0$, and $R_0$ are standard representations as given in first 230 groups of this listing, i.e. the category 1) groups (also these standard representations of the conventional space groups in the International Tables for Crystallography). Since $Q_0 = Q$ for all groups in this listing, the first matrix is always an identity matrix. If a matrix is not applicable to the current category then zeros are given for all elements of the matrix.

The fourth line contains between three and six numbers. These give the color of the translation subgroup generators; 1 means colorless (coupled with 1), 2 means primed (colored with 1'), 3 means starred (colored with 1*), and 4 means prime-starred (colored with 1'*). The translation indicated by each position is depends on the lattice type of $Q$ as follows:

|   | 1st position | 2nd position | 3rd position | 4th position | 5th position | 6th position |
|---|---|---|---|---|---|---|
| P | $t_{[100]}$ | $t_{[010]}$ | $t_{[001]}$ | | | |
| C | $t_{[100]}$ | $t_{[010]}$ | $t_{[001]}$ | $t_{[½½0]}$ | | |
| A | $t_{[100]}$ | $t_{[010]}$ | $t_{[001]}$ | $t_{[0½½]}$ | | |
| I | $t_{[100]}$ | $t_{[010]}$ | $t_{[001]}$ | $t_{[½½½]}$ | | |
| F | $t_{[100]}$ | $t_{[010]}$ | $t_{[001]}$ | $t_{[0½½]}$ | $t_{[½0½]}$ | $t_{[½½0]}$ |

| R | $t_{[100]}$ | $t_{[010]}$ | $t_{[001]}$ | $t_{[⅔⅓⅓]}$ | $t_{[⅓⅔⅔]}$ | |

The 4$^{th}$ through 6$^{th}$ positions of the fourth line are always centering translations (those with non-integer values). It is only necessary to use the first three positions generally because the coloring of the centering translations is also given in the fifth through eighth lines.

The fifth line contains a list of numbers whose length is a multiple of 16. This list is to be partitioned into 4-by-4 matrices representing the matrix form of the colorless operations. The sixth, seventh, and eighth lines are to be similarly partitioned and represent the matrix form of primed, starred and prime-starred operations respectively.

### Appendix E. The automorphism group of 1'1* and the subgroups thereof

**1'1*** is isomorphic to $\mathbb{Z}_2 \times \mathbb{Z}_2$. We can use this isomorphism to derive Aut(**1'1***). Let $m$ be the isomorphism of **1'1*** onto $\mathbb{Z}_2 \times \mathbb{Z}_2$ which maps 1 to (0,0), 1' to (1,0), 1* to (0,1), and 1'* to (1,1). Using this isomorphism, the automorphism group of **1'1*** can be solved for from the automorphism group of $\mathbb{Z}_2 \times \mathbb{Z}_2$:

**Equation 16**

$$\text{Aut}(\mathbf{1'1^*}) = m\,\text{Aut}(\mathbb{Z}_2 \times \mathbb{Z}_2)\,m^{-1}$$

The automorphism group of $\mathbb{Z}_2 \times \mathbb{Z}_2$ is GL(2,2), the general linear group of degree 2 over the field of 2 elements, or more precisely is isomorphic to it. GL(2,2) is represented by these six matrices:

$\{\begin{pmatrix}1 & 0\\0 & 1\end{pmatrix}, \begin{pmatrix}1 & 1\\1 & 0\end{pmatrix}, \begin{pmatrix}0 & 1\\1 & 1\end{pmatrix}, \begin{pmatrix}0 & 1\\1 & 0\end{pmatrix}, \begin{pmatrix}1 & 1\\0 & 1\end{pmatrix}, \begin{pmatrix}1 & 0\\1 & 1\end{pmatrix}\}$. Using $m$, we can solve for the **1'1*** automorphism implied by each of these matrices. As an example, consider the **1'1*** automorphism, $m\begin{pmatrix}1 & 1\\1 & 0\end{pmatrix}m^{-1}$, implied by the $\mathbb{Z}_2 \times \mathbb{Z}_2$ automorphism $\begin{pmatrix}1 & 1\\1 & 0\end{pmatrix}$:

$$(0,0) \xrightarrow{\begin{pmatrix}1 & 1\\1 & 0\end{pmatrix}} (0,0) \leftrightarrow 1 \xrightarrow{m\begin{pmatrix}1 & 1\\1 & 0\end{pmatrix}m^{-1}} 1$$

$$(1,0) \xrightarrow{\begin{pmatrix}1 & 1\\1 & 0\end{pmatrix}} (1,1) \leftrightarrow 1' \xrightarrow{m\begin{pmatrix}1 & 1\\1 & 0\end{pmatrix}m^{-1}} 1'^*$$

$$(0,1) \xrightarrow{\begin{pmatrix}1 & 1\\1 & 0\end{pmatrix}} (1,0) \leftrightarrow 1^* \xrightarrow{m\begin{pmatrix}1 & 1\\1 & 0\end{pmatrix}m^{-1}} 1'$$

$$(1,1) \xrightarrow{\begin{pmatrix}1 & 1\\1 & 0\end{pmatrix}} (0,1) \leftrightarrow 1'^* \xrightarrow{m\begin{pmatrix}1 & 1\\1 & 0\end{pmatrix}m^{-1}} 1^*$$

Thus, $m \begin{pmatrix} 1 & 1 \\ 1 & 0 \end{pmatrix} m^{-1}$ is the permutation $\begin{pmatrix} 1' & 1^* & 1'^* \\ 1'^* & 1' & 1^* \end{pmatrix}$. Or using one-line notation, (1'*, 1', 1*).

Applying this to each element of GL(2,2), we find that $\begin{pmatrix} 1 & 0 \\ 0 & 1 \end{pmatrix}, \begin{pmatrix} 1 & 1 \\ 1 & 0 \end{pmatrix}, \begin{pmatrix} 0 & 1 \\ 1 & 1 \end{pmatrix}, \begin{pmatrix} 0 & 1 \\ 1 & 0 \end{pmatrix}, \begin{pmatrix} 1 & 1 \\ 0 & 1 \end{pmatrix}$, and $\begin{pmatrix} 1 & 0 \\ 1 & 1 \end{pmatrix}$ imply (1', 1*, 1'*), (1'*, 1', 1*), (1*, 1'*,1'), (1*, 1', 1'*), (1', 1'*, 1*), and (1'*, 1*, 1') respectively.

The subgroups of GL(2,2) can be used to find the subgroups of Aut(**1'1***). The subgroups of GL(2,2):

Index=6 (order = 1): $\{\begin{pmatrix} 1 & 0 \\ 0 & 1 \end{pmatrix}\}$

Index=3 (order=2): $\{\begin{pmatrix} 1 & 0 \\ 0 & 1 \end{pmatrix}, \begin{pmatrix} 0 & 1 \\ 1 & 0 \end{pmatrix}\}, \{\begin{pmatrix} 1 & 0 \\ 0 & 1 \end{pmatrix}, \begin{pmatrix} 1 & 1 \\ 0 & 1 \end{pmatrix}\}$, and $\{\begin{pmatrix} 1 & 0 \\ 0 & 1 \end{pmatrix}, \begin{pmatrix} 1 & 0 \\ 1 & 1 \end{pmatrix}\}$

Index=2 (order=3): $\{\begin{pmatrix} 1 & 0 \\ 0 & 1 \end{pmatrix}, \begin{pmatrix} 1 & 1 \\ 1 & 0 \end{pmatrix}, \begin{pmatrix} 0 & 1 \\ 1 & 1 \end{pmatrix}\}$

Index=1 (order=6): $\{\begin{pmatrix} 1 & 0 \\ 0 & 1 \end{pmatrix}, \begin{pmatrix} 1 & 1 \\ 1 & 0 \end{pmatrix}, \begin{pmatrix} 0 & 1 \\ 1 & 1 \end{pmatrix}, \begin{pmatrix} 0 & 1 \\ 1 & 0 \end{pmatrix}, \begin{pmatrix} 1 & 1 \\ 0 & 1 \end{pmatrix}, \begin{pmatrix} 1 & 0 \\ 1 & 1 \end{pmatrix}\}$

imply the following Aut(**1'1***) subgroups:

Index=6 (order = 1): {(1', 1*, 1'*)}

Index=3 (order=2): {(1', 1*, 1'*), (1', 1'*, 1*)}, {(1', 1*, 1'*), (1*, 1', 1'*)}, and {(1', 1*, 1'*), (1'*, 1*, 1')}

Index=2 (order=3): {(1', 1*, 1'*), (1*, 1'*, 1'), (1'*, 1', 1*)}

Index=1 (order=6): {(1', 1*, 1'*), (1', 1'*, 1*), (1*, 1', 1'*), (1*, 1'*, 1'), (1'*, 1', 1*), (1'*, 1*, 1')}

For each line of generators, Zamorzaev & Palistrant (Zamorzaev & Palistrant, 1980) give the number of types represented by that line, which corresponds to the order of a subgroup of Aut(**1'1***). Zamorzaev & Palistrant do not give which automorphisms need to be applied to generate these types. However, any valid partition of Aut(**1'1***) by generated type must have an associated subgroup by which the elements of the members of the partition must be related, i.e. said partition is a coset decomposition of Aut(**1'1***).

Applying this understanding, we find that for lines of generators which result in 1, 2, or 6 types, there is only one valid partition of Aut(**1'1***) by generated type because there is only one coset decomposition for index 1, 2, and 6 respectively. For lines of generators which result in 3 types, there are three possible coset decompositions (technically three left and three right, but we can only choose to use left or right multiplication on the generating set, not both).

The results from this analysis for a line of generators can be summarized as follows:

| (1', 1*, 1'*) | (1*, 1'*, 1') | (1'*, 1', 1*) |
|---|---|---|
| (1', 1'*, 1*) | (1*, 1', 1'*) | (1'*, 1*, 1') |

1 type = all permutations yield groups of the same type

2 types = permutations in the same row yield groups of the same type

3 types = permutations in the same row yield distinct types of groups

6 types = all permutations are distinct types of groups

This method can be generalized to multiple antisymmetry and other color symmetries. For double antisymmetry, it would have been simpler to check which permutations of anti-identities preserved the group structure of **1'1***. As it turns out, they all do, i.e. Aut(**1′1***) $\cong$ **S$_3$**. This will not work $l$-multiple antisymmetry for $l > 2$. For multiple antisymmetry: the automorphism group of $\mathbb{Z}_2^n$ is isomorphic to general linear group GL(n,2) of degree $n$ over the field of *2* elements and a similar method can thus be employed.

**Figure 1:** Identity (1) and anti-identities (1', 1*, and 1'*) of the rotation-reversal and time-reversal space groups.

|  | Does not reverse time | Reverses time |
|---|---|---|
| **Does not reverse rotation** | 1 | 1' |
| **Reverses rotation** | 1* | 1'* |

**Figure 2: a)** <u>Multiplication table of **1'1\***.</u> To evaluate the product of two elements, we find the row associated with the first element and the column associated with second element, e.g. for 1'·1*, go to the second row third column to find 1'*. **b)** <u>Cayley graph generated by 1', 1*, and 1'*.</u> To evaluate the product of two elements, we start from the circle representing the first element and follow the arrow representing the second, e.g. for 1'·1*, we start on the red circle (1') and take the blue path (1*) to the green circle (1'*).

a)

|  | 1 | 1' | 1* | 1'* |
|---|---|---|---|---|
| 1 | 1 | 1' | 1* | 1'* |
| 1' | 1' | 1 | 1'* | 1* |
| 1* | 1* | 1'* | 1 | 1' |
| 1'* | 1'* | 1* | 1' | 1 |

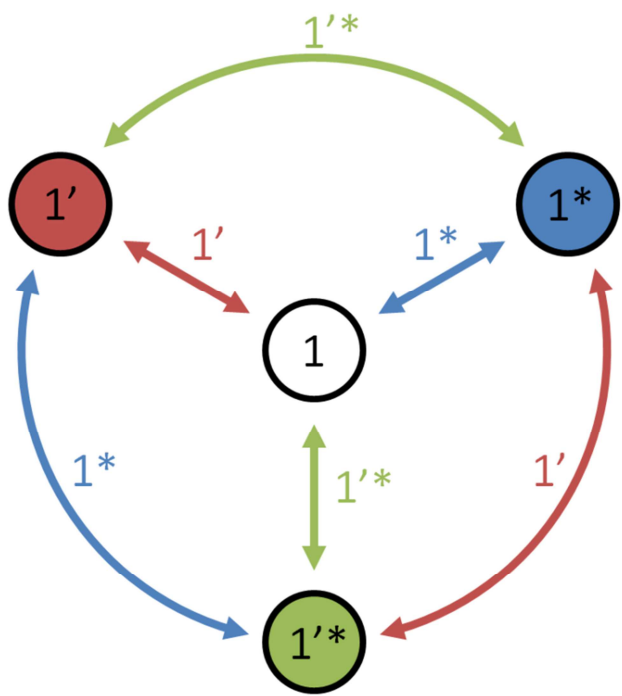

**Figure 3:** Demonstration of proper affine equivalence of **Q(H){K}** groups generated for **Q = 222** using point group diagrams (stereoscopic projections).

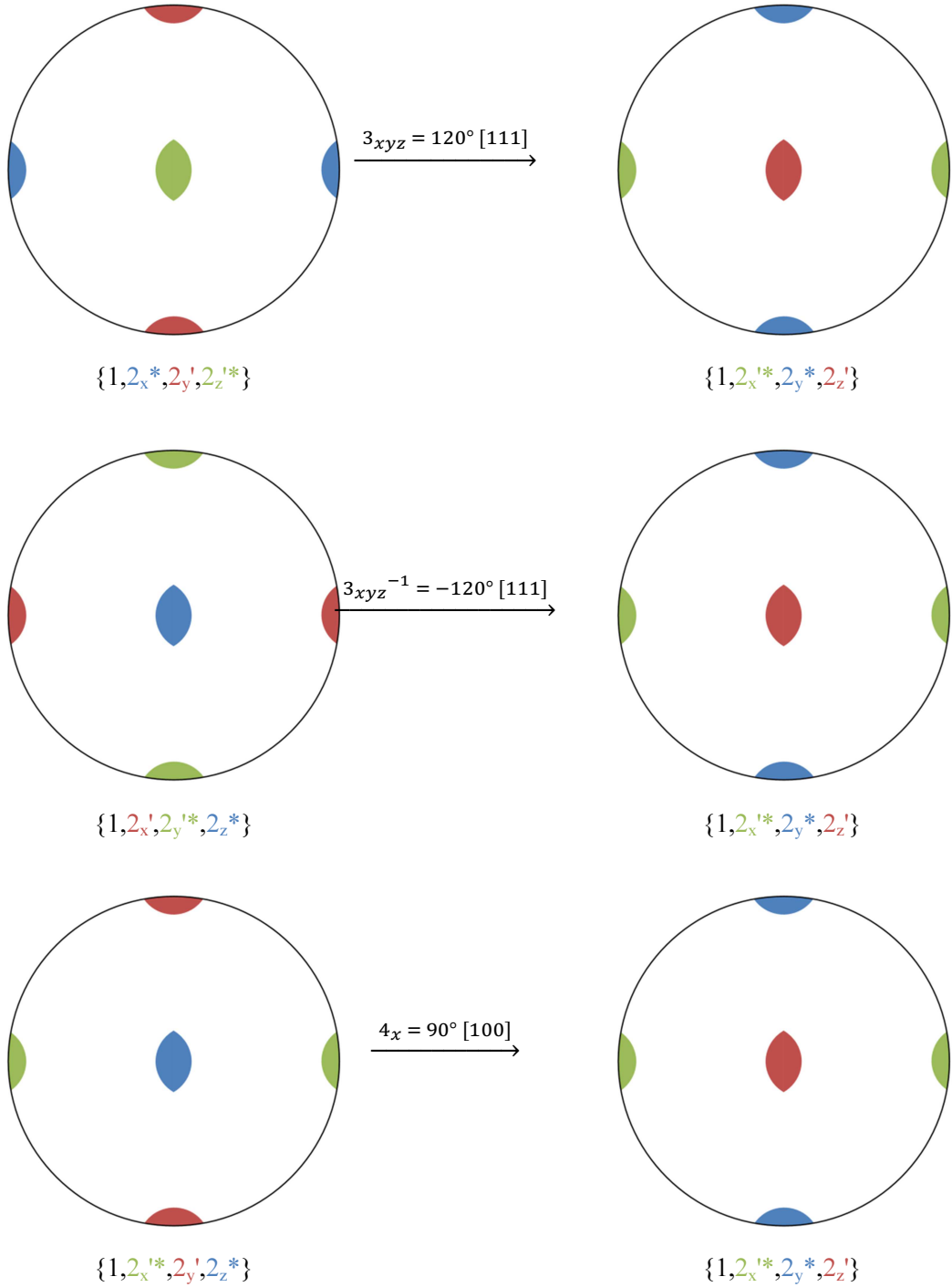

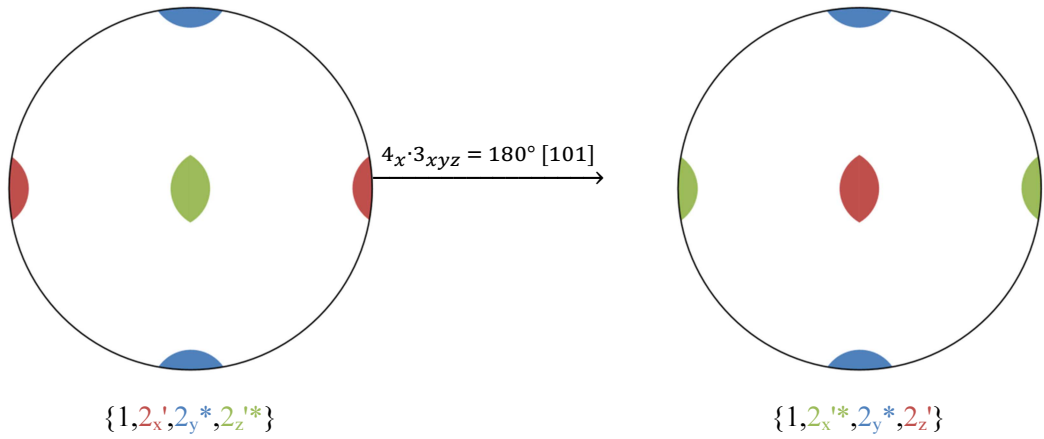

$\{1, 2_x', 2_y^*, 2_z'^*\}$ → $4_x \cdot 3_{xyz} = 180°\ [101]$ → $\{1, 2_x'^*, 2_y^*, 2_z'\}$

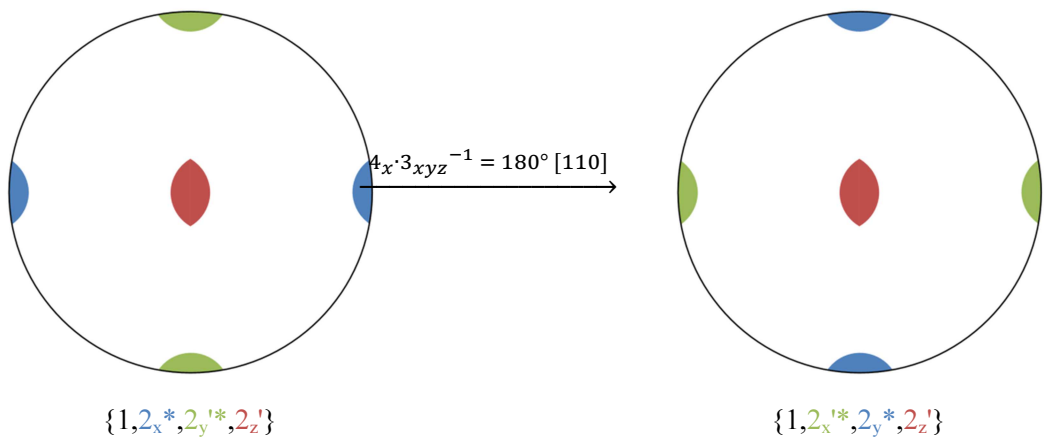

$\{1, 2_x^*, 2_y'^*, 2_z'\}$ → $4_x \cdot 3_{xyz}^{-1} = 180°\ [110]$ → $\{1, 2_x'^*, 2_y^*, 2_z'\}$

**Figure 4:** Example double antisymmetry space group diagrams.

**a)** No. 8543, C2'/m*

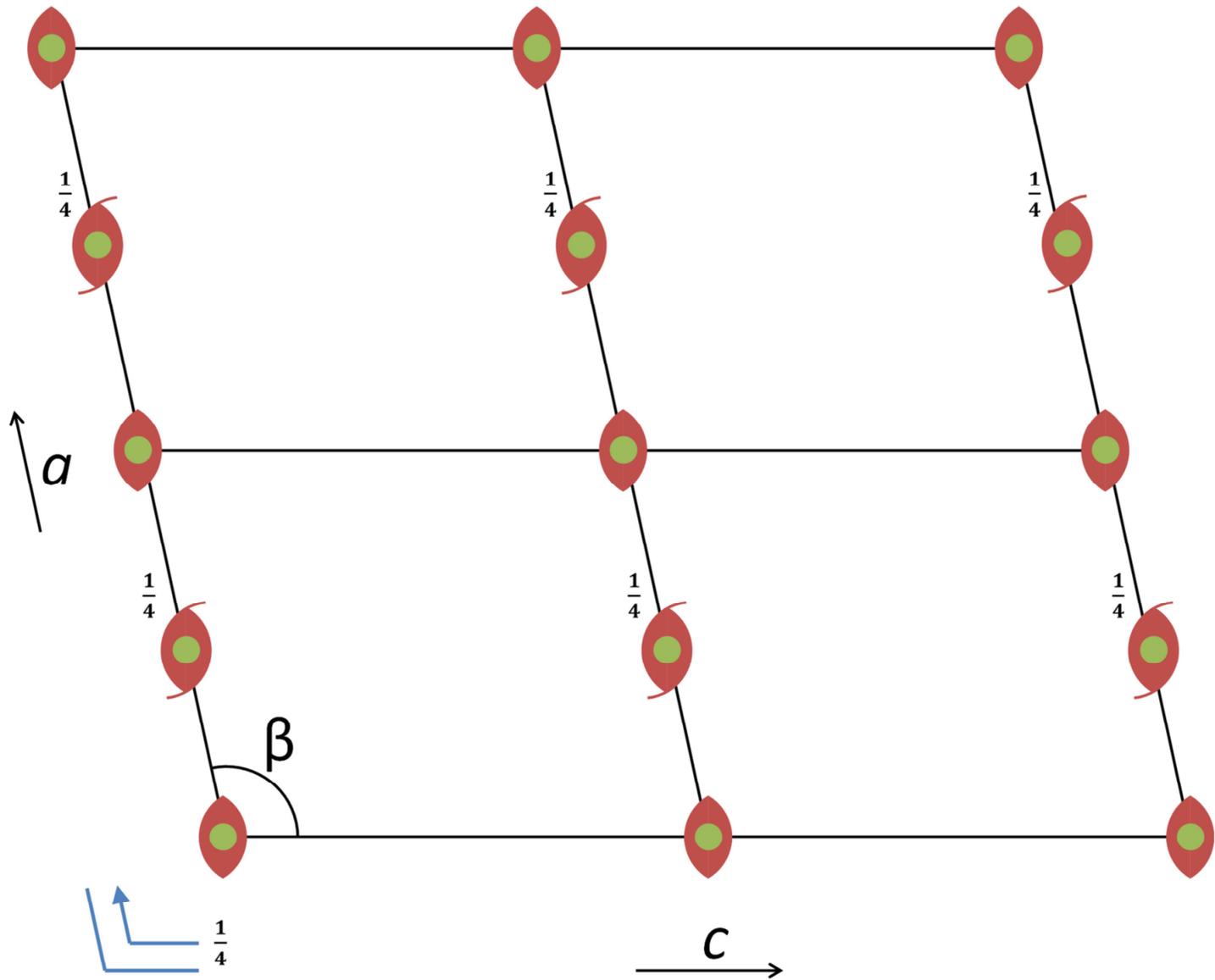

**b)** No. 16490, I4*/mm'm'*

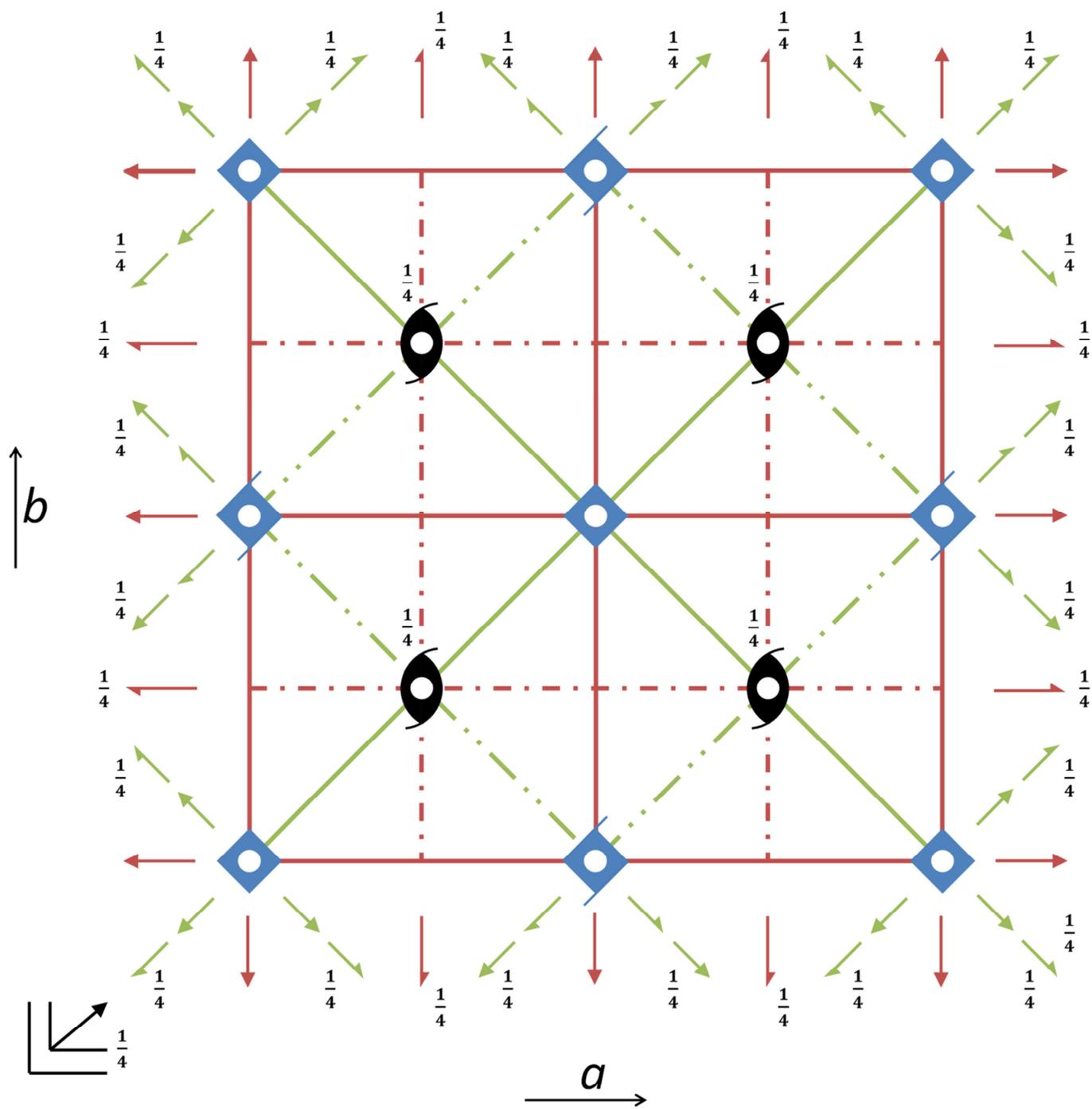

**c)** No. 13461, Ib'*c'a'

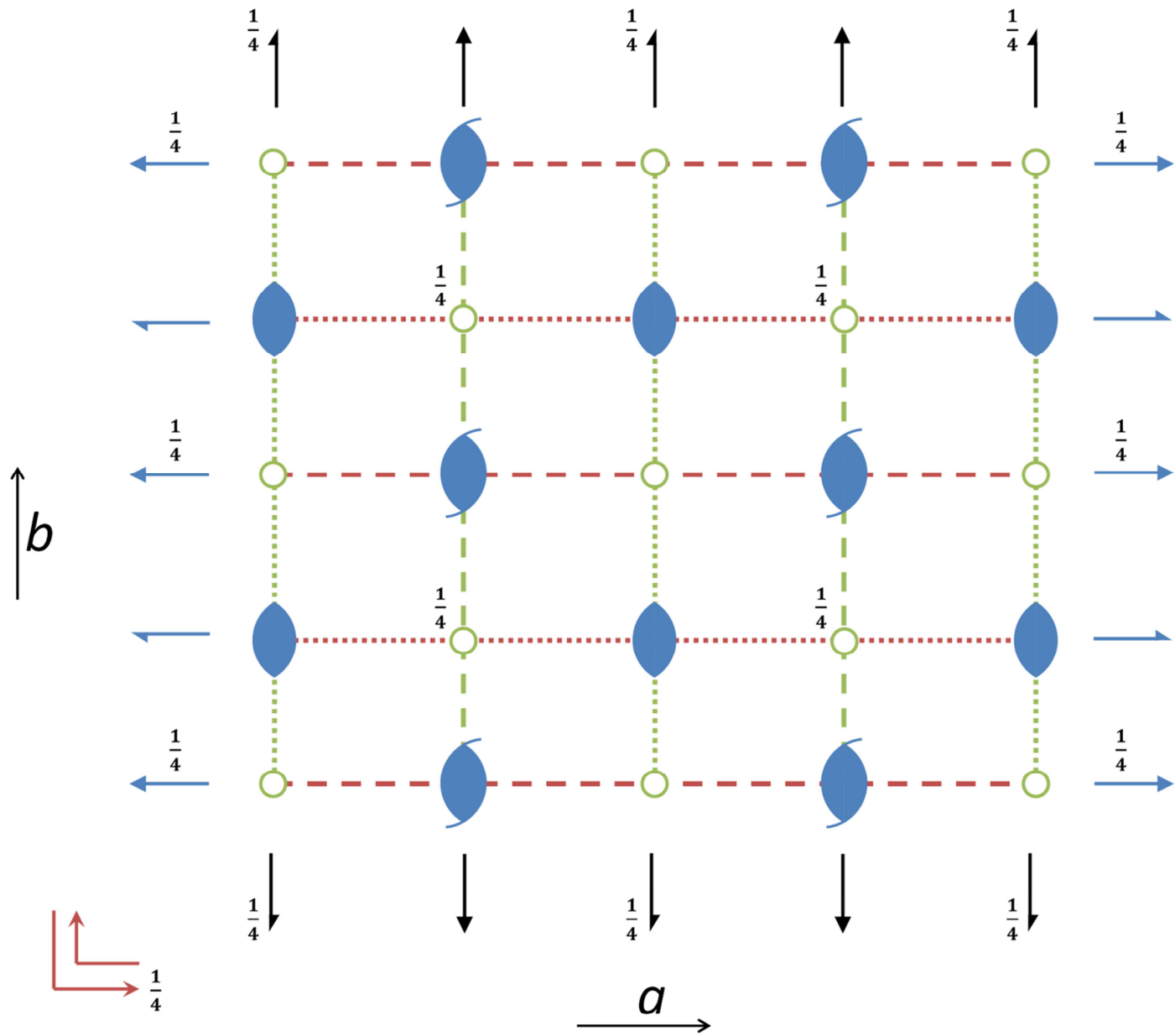